\documentclass[letterpaper,11pt]{article}
\usepackage{amsmath, epsfig, amssymb,array}

\begin{document}

\newcommand{\bref}[1]{(\ref{#1})}
\def\sl#1{\mathord{\not\mathrel{{\mathrel{#1}}}}}
\def\ol#1{\overline{#1}}
\baselineskip=18pt


\thispagestyle{empty}
\vspace{20pt}
\font\cmss=cmss10 \font\cmsss=cmss10 at 7pt

\begin{flushright}
\today \\
UMD-PP-08-015 \\
\end{flushright}

\hfill
\vspace{20pt}

\begin{center}
{\Large \textbf {Flavor Violation Tests of Warped/Composite SM in
the Two-Site Approach } }
\end{center}

\vspace{15pt}

\begin{center}
{\large Kaustubh Agashe$\, ^{}$,
Aleksandr Azatov$\, ^{}$ and Lijun Zhu$\, ^{}$} \\
\vspace{15pt} $^{}$\textit{Maryland Center for Fundamental Physics,
     Department of Physics,
     University of Maryland,
     College Park, MD 20742, USA}

\end{center}

\vspace{20pt}

\begin{center}
\textbf{Abstract}
\end{center}
\vspace{5pt} {\small \noindent

We study flavor violation in the quark sector in a purely $4D$,
two-site effective field theory description of the Standard Model
and just their first Kaluza-Klein excitations from a warped extra
dimension. The warped $5D$ framework can provide solutions to both
the Planck-weak and flavor hierarchies of the SM. It is also related
(via the AdS/CFT correspondence) to partial compositeness of the SM.
We focus on the dominant contributions in the two-site model to two
observables which we argue provide the strongest constraints from
flavor violation,
namely, $\epsilon_K$ and BR $\left( b \rightarrow s \gamma \right)$,
where contributions in the two-site model occur at tree and
loop-level, respectively. In particular, we demonstrate that a
``tension'' exists between these two observables in the sense that
they have opposite dependence on composite site Yukawa couplings,
making it difficult to decouple flavor-violating effects using this
parameter. We choose the size of the composite site QCD coupling
based on the relation of the two-site model to the $5D$ model
(addressing the Planck-weak hierarchy), where we match the $5D$ QCD
coupling to the $4D$ coupling at the loop-level and assuming
negligible tree-level brane-localized kinetic terms. We estimate
that a larger size of the $5D$ gauge coupling is constrained by the
requirement of $5D$ perturbativity. We find that $\sim O(5)$ TeV
mass scale for the new particles in the two-site model can then be
consistent with both observables. We also compare our analysis of
$\epsilon_K$ in the two-site model to that in $5D$ models, including
both the cases of a brane-localized and bulk Higgs.

}

\vfill\eject \noindent


\section{Introduction}

The framework of a warped extra dimension with Standard Model (SM)
fields propagating in the bulk \cite{bulk, gn, gp} is a very
attractive extension of the SM since it can provide solutions to
both the Planck-weak \cite{rs1} and flavor hierarchy problems of the
SM \cite{gn, gp}.
Moreover, the versions of this framework with a grand unified gauge
symmetry in the bulk can naturally lead to precision unification of
the three SM gauge couplings \cite{Agashe:2005vg} and
%
%
a candidate for the dark matter of the universe (the latter from
requiring longevity of the proton) \cite{Agashe:2004ci}.
The new particles in this framework are the Kaluza-Klein (KK)
excitations of the SM fields with mass at the $\sim$ TeV scale. Such a
framework can thus give significant contributions to various
precision tests of the SM.
The electroweak precision tests (EWPT) can be satisfied for KK mass
scale of a few TeV \cite{Agashe:2003zs,
Agashe:2004rs, EWPTmodel} using suitable custodial
symmetries \cite{Agashe:2003zs, Agashe:2006at}. 

In this paper, we focus on the
solution to the flavor hierarchy of the SM in the framework of warped
extra dimension
and the resulting
flavor-violation. 
The idea is
that the effective $4D$ Yukawa couplings of the SM fermions are given
by a product
%
%
of the fundamental $5D$ Yukawa couplings and
the overlap of the profiles (of the SM fermions and 
the Higgs) in the extra dimension.
Moreover, vastly different
profiles in the extra dimension for the SM fermions (which are the zero-modes
of $5D$ fermions), and hence their hierarchical
overlaps with Higgs,
can be easily obtained by small variations
in the $5D$ fermion mass parameters.
Thus, hierarchies in the $4D$ Yukawa couplings
can be generated without any (large) hierarchies in
the fundamental $5D$ parameters ($5D$ Yukawa couplings
and $5D$ mass parameters for fermions).
As a corollary, the couplings of SM fermions (with
different profiles) to KK
modes (again following from the relevant overlaps of profiles)
are non-universal, resulting in flavor violation
from exchange of these KK modes \cite{Delgado:1999sv}.
However, there is a built-in analog of GIM mechanism of the SM in
this framework \cite{gp,hs, Agashe:2004cp} which suppresses flavor
changing neutral currents (FCNC's). Namely, the non-universalities
in couplings of SM fermions to KK modes are of the size of $4D$ Yukawa
couplings since KK modes have a similar profile to the SM Higgs.

In spite of this analog of the GIM mechanism, it was shown recently
\cite{Csaki:2008zd, Blanke:2008zb} (see also
\cite{Fitzpatrick:2007sa, Davidson:2007si}) that the constraint on
KK mass scale from contributions of KK gluon to $\epsilon_K$ is
quite stringent.
In particular, for the model with the
SM Higgs (strictly) localized on the TeV brane in a $5D$ slice
of anti-de Sitter space (AdS),
the limit on the KK
mass scale from $\epsilon_K$ is $\sim 10$ TeV
%
%
for the smallest allowed $5D$ QCD coupling
obtained by {\em loop}-level matching to the $4D$ coupling with negligible
tree-level brane kinetic terms (in the framework which addresses the
Planck-weak hierarchy).
On the other hand, for larger brane kinetic terms such
that the $5D$ QCD coupling (in units of
the AdS curvature scale) is
$\sim 4 \pi$, the lower limit on KK mass scale increases to $\sim 40$ TeV.
In addition, the constraint on the KK mass scale
is weakened as the size of the $5D$ Yukawa (in units of the
AdS curvature scale) is increased.
However,
this direction reduces the regime of validity of the
$5D$ effective field theory (EFT):
the above limits on KK mass scale are for the size of $5D$ Yukawa 
such that about two KK modes are allowed in the $5D$ EFT.

In the light of these constraints, instead of an ``anarchic''
approach to flavor in $5D$, i.e.,
no hierarchies or relations in the various $5D$ flavor
parameters,
references \cite{Fitzpatrick:2007sa,
Cacciapaglia:2007fw, Csaki:2008eh} have proposed imposing $5D$
flavor symmetries in order to relate these
parameters and hence to suppress flavor violation.
Lower KK mass scales are thereby allowed, improving the fine-tuning in
electroweak symmetry breaking (EWSB) and the discovery potential at the
Large Hadron Collider (LHC).
For other flavor
studies in this framework, see references \cite{others1, others2,
Agashe:2004ay, Agashe:2006iy, Casagrande:2008hr}.

However, the phenomenology of the TeV-scale KK modes and the SM
Higgs is quite sensitive to the structure near the TeV brane (where
these particles are localized). For example, the SM Higgs can be the
lightest mode of a $5D$ scalar (instead of being a strictly TeV
brane-localized field), but with a profile which is still peaked
near the TeV brane (such that the Planck-weak hierarchy is still
addressed) -- we will denote this scenario by ``bulk Higgs''
\cite{Davoudiasl:2005uu, Cacciapaglia:2006mz}.\footnote{Note that
the models where Higgs is the $5^{ \hbox{th } }$ component of a $5D$
gauge field \cite{Contino:2003ve, Agashe:2004rs} do {\em not} belong
to this class.}
Moreover, the warped
geometry might deviate from pure AdS near the TeV brane which in
fact could be replaced with a ``soft wall'' \cite{soft}. Similarly,
in general, there are non-zero TeV brane-localized kinetic terms for
the bulk fields \cite{Davoudiasl:2002ua}.
Such variations of the minimal models are not likely to modify the
couplings and spectrum of the KK modes/Higgs significantly -- for example,
the constraint on KK mass scale from various precision tests will
not be modified by much more than $O(1)$ factors. However, even such
modest changes can dramatically impact the LHC signals, especially
the production cross-sections for the KK modes.

Instead of focusing on a {\em specific} limit of the full $5D$
model, such considerations then strongly motivate analyzing the
phenomenology of this framework using a more economical description
(for example, using fewer parameters than the $5D$ models)
which can capture its
%
%
robust aspects.
%
%
Such an approach is
provided by the ``two site model'' \cite{Contino:2006nn} which is a
purely $4D$ effective field theory obtained by truncating the $5D$
AdS model to the SM particles and their first KK excitations,
roughly achieved by ``deconstruction/discretization''
\cite{ArkaniHamed:2001ca} of the warped extra dimension.
Equivalently, based on the AdS/CFT correspondence
\cite{Maldacena:1997re} as applied to a slice of AdS
\cite{Arkani-Hamed:2000ds}, the two-site model also describes two
sectors: composites of purely $4D$ strong dynamics and elementary
fields (which are not part of the strong dynamics). These two
sectors mix, with the resulting mass eigenstates being the SM
particles and their heavier partners, which correspond to the zero and KK
modes of the $5D$ model.
With this motivation in mind, an analysis of EWPT in two-site model
was performed in \cite{Contino:2006nn}.

In this paper, we study flavor-violation in the quark sector in this
two-site model, in the incarnation {\em corresponding to flavor
anarchy in the $5D$ AdS theory}.
We focus on effects of tree-level heavy gluon exchange in
$\epsilon_K$ and Higgs-heavy fermion loops in BR $\left( b
\rightarrow s \gamma \right)$\footnote{{\em Estimates} for $b
\rightarrow s \gamma$ in the $5D$ AdS model were performed in
references \cite{Agashe:2004cp, Agashe:2004ay}.}. We will show that
a {\em combination} of
these processes provide the strongest constraints on the two-site
model (and hence probably on the general framework of warped extra
dimension).
We leave a more complete study of flavor violation in the two-site
model, including other contributions to these observables and a
global analysis (i.e., including other observables), for future
work.

A central observation of our analysis is that
\begin{itemize}

\item
a ``tension'' exists between the two observables $\epsilon_K$ and BR
$\left( b \rightarrow s \gamma \right)$ in the sense that they have
opposite dependence on the composite site Yukawa coupling so that it
is {\em not} possible to simultaneously suppress both these flavor
violating effects using this parameter\footnote{A similar effect was
found earlier with regard to the $5D$ Yukawa coupling during an
analysis of lepton flavor violation in the $5D$ AdS model
\cite{Agashe:2006iy}}.

\end{itemize}
However, $\epsilon_K$ can be suppressed by choosing small composite
site QCD coupling. We find that
\begin{itemize}

\item
$\sim O(5)$ TeV mass scale for the heavy states is allowed
simultaneously by $\epsilon_K$ and BR $\left( b \rightarrow s \gamma
\right)$ for a size of the composite site QCD gauge coupling
corresponding to loop-level matching of the $5D$ QCD coupling
(with no tree-level brane kinetic terms) to the $4D$ coupling
in the $5D$ model which addresses the
%
%
Planck-weak hierarchy.
%
%

\end{itemize}

In fact, we argue that,
\begin{itemize}

\item
once we include color factors
in the {\em estimate} of the loop expansion parameter,
a
$5D$ QCD coupling larger than the above value
%
%
might lead to the $5D$ theory no longer being perturbative.
%
%

\end{itemize}

Note that, even with the above smallest value of the $5D$ QCD coupling, the
lower limit on mass scale of new particles is different for
the two-site model ($\sim O(5)$ TeV) as compared to the $5D$ model with
brane-localized Higgs ($\sim 10$ TeV). This is partly due to the fact that,
in spite of the two-site model being a deconstruction of the $5D$ model, the
detailed features of the two models are different. Secondly, the
above bound for the two-site model is from a {\em combination} of
$\epsilon_K$ and BR $\left( b \rightarrow s \gamma \right)$, whereas
that for the brane-localized Higgs model is from $\epsilon_K$ only.

In fact, we compare in detail
our results for the two-site model to those in $5D$ AdS
models. In particular, we find that
\begin{itemize}

\item
the {\em relations} between the couplings of various particles
%
%
(at least the ones relevant to
$\epsilon_K$) in the two-site model ``mimic'' those
between the corresponding couplings
%
%
in models with
{\em bulk} Higgs,
instead of the case of brane-localized Higgs which has been analyzed
in the literature thus far.

\end{itemize}

So, our bounds from $\epsilon_K$ for the two-site model apply
directly to the $5D$ AdS models with bulk Higgs
(of a specific profile) for the choice of
composite gauge and Yukawa site couplings being same as the
corresponding purely KK couplings in the $5D$ model. 
And, 
we expect
a tension between $\epsilon_K $ and $b \rightarrow s \gamma$
in the $5D$ AdS model (similar
to that in the two-site model). Thus our analysis for the two-site model
suggests that a KK scale
as low as $\sim O(5)$ TeV might also be allowed in the $5D$ AdS
models with bulk Higgs by the {\em combined} constraints from $\epsilon_K$
and BR $\left( b \rightarrow s \gamma \right)$.
On the other hand, we show that
\begin{itemize}

\item
if, instead of using $b \rightarrow s \gamma$ to
place an upper bound on the $5D$ Yukawa,
we restrict it only by
the requirement that two KK modes in the $5D$ EFT are allowed,
then a KK mass scale as low as $\sim O(3)$ TeV
might be consistent with the constraints
from $\epsilon_K$ (only) in the $5D$ model with {\em bulk} Higgs.

\end{itemize}

The outline of the rest of our paper is as follows. We begin with a
review of the relevant features of two-site model, especially the
couplings which will be used in our analysis of flavor constraints
on this model. Assuming anarchic composite site Yukawa couplings,
one typically finds multiple terms (of similar size) in the
flavor-violating amplitudes, with $O(1)$ free parameters in the
mixing angles and phases so that we need to scan over these
parameters. It is then useful to present {\em analytical} formulae for {\em one}
such generic contribution (with mixing angles set to their
``natural'' size) in $\epsilon_K$ and $b \rightarrow s \gamma$. This
exercise is presented in sections \ref{ek} and \ref{bsg}, providing
an {\em estimate} of the bounds.
In section \ref{Zbb}, we briefly discuss the bound from $Z b
\bar{b}$, which (although not flavor-violating) turns out to be
relevant for the analysis of flavor violation.
The results of the {\em numerical} analysis which includes the full
amplitudes (summing up all terms) for $\epsilon_K$ and $b
\rightarrow s \gamma$ (and $Z b \bar{b}$) are presented in section
\ref{numerical}.
The
allowed value of
$\sim O(5)$ TeV
for mass scale of
heavy particles
in the two-site model mentioned above is
based on a
combination of the numerical analysis and the analytical estimates.
We conclude in section \ref{conclusion}. Several
appendices deal with further aspects of our analysis. In particular,
here we briefly discuss other contributions to $\epsilon_K$ and
other $B$-physics observables which provide weaker constraints on
the two-site model. We present details of the loop calculation for
$b \rightarrow s \gamma$ and the exact (numerical) scanning
procedure that we used. In a final appendix, we contrast our results
for the two-site model with those for the $5D$ AdS model. This
comparison is summarized in Table \ref{table}.

\section{Review of Two-Site Model}

\subsection{Elementary and Composite Sectors}
\label{review}

In this section, we review the basic features of the two-site model (for
more details see \cite{Contino:2006nn}).  The particle
content is divided into two sectors: composite and elementary. The elementary
sector of the model is equal exactly to that of SM except for
the Higgs field. The SM gauge fields ($SU(3)\otimes SU(2)_L\otimes
U(1)_Y$) will be denoted in the following way,
\begin{eqnarray}
A_\mu\equiv \{G_\mu,W_\mu,B_\mu\}
\end{eqnarray}
and fermion $SU(2)_L$ doublets by,
\begin{eqnarray}
\psi_L\equiv \{ q_{Li}=(u_{Li},d_{Li}),l_{Li}=(\nu_{Li},e_{Li})\}
\end{eqnarray}
and finally $SU(2)_L$ singlets as,
\begin{eqnarray}
\tilde\psi_R\equiv \{ u_{Ri},d_{Ri},\nu_{Ri},e_{Ri}\}.
\end{eqnarray}
The only renormalizable interactions are  the gauge interactions.
\begin{eqnarray}
\mathcal{
L}^{\small{\text{elementary}}}=-\frac{1}{4}F_{\mu\nu}^2+\bar{\psi}_L
i \sl{D} \psi_L+\bar{\tilde \psi}_R i \sl D \tilde \psi_R .
\end{eqnarray}

The composite boson sector (containing the
SM Higgs and massive spin $1$ particles) has $SU(3)\otimes
SU(2)_L\otimes SU(2)_R\otimes U(1)_X$ global symmetries, where we need
the additional custodial $SU(2)_R$ to suppress new physics
contribution to the
%
%
$T$ parameter
\cite{Agashe:2003zs}. There are fifteen heavy vector mesons
($\rho_\mu$) that belong to adjoint representation of the
$SU(3)\otimes SU(2)_L\otimes SU(2)_R\otimes U(1)_X$, and they can
be decomposed into two sets: $\rho_*$, which are in the adjoint
representation of the SM gauge group  and their orthogonal
combinations $\tilde \rho$
\begin{equation}
\label{rho mesons} \rho^*_\mu = \{ G_\mu^* , W_\mu^* ,{\cal
B}_\mu^* \}\, , \qquad
 \tilde\rho_\mu = \Big\{ \tilde W^{\pm}_\mu \equiv \frac{\tilde W_1 \mp i\, \tilde W_2}{\sqrt{2}} ,
 \tilde {\cal B}_\mu \Big\} \, .
\end{equation}
We associate ${\cal B^*},{\cal \tilde B}$ with the  generators
$T_{{\cal B}^*}=Y_{\text{hypercharge}}=\frac{T^{3R}+\sqrt{2/3}
T_X}{\sqrt{5/3}}$  and $T_{{\cal \tilde B}}=\frac{T^{3R}-\sqrt{2/3}
T_X}{\sqrt{5/3}}$,  where $T_{{\cal B}^*}$ is hypercharge generator
in the SO(10) normalization. Higgs field belongs to the composite
sector and  is a real bidoublet under $SU(2)_L\otimes SU(2)_R$:
$(H,\tilde H)$.

Every SM fermion representation will be accompanied  by a heavy
composite Dirac fermion, so the composite sector will consist of
$SU(2)_L$ doublets :
\begin{eqnarray}
\chi\equiv\left(Q_i=\{U_i,D_i\},L_i=\{N_i,E_i \}\right)
\end{eqnarray}
and $SU(2)_L$ singlets:
\begin{eqnarray}
\tilde\chi=\left(\tilde U_i,\tilde D_i,\tilde E_i,\tilde N_i\right)
\end{eqnarray}
They are all singlets under $SU(2)_R$. The Dirac masses of the
composite sector doublets and singlets are $m_*,\tilde
m_*$, respectively, which we assume to be the same (and generation-independent)
for simplicity. $U(1)_X$ charges for fermions
are chosen to reproduce the usual SM hypercharges.

The Lagrangian of the composite sector is
\begin{eqnarray}
\label{complagr}
\mathcal{L}_{\small{\text{composite}}}=-\frac{1}{4}\rho_{\mu\nu}^2+\frac{M_{*}^2}{2}\rho_\mu^2+|D_\mu H|^2-V(H)+\nonumber\\
+\bar\chi(i\sl D -m_*)\chi+\bar{\tilde \chi}(i\sl D -\tilde
m_*)\tilde \chi-\bar\chi(Y^{u}_{*}\tilde H \tilde\chi^u+Y^d_{*} H
\tilde\chi^d)+h.c.
\end{eqnarray}
where $M_*$ is the mass of the composite sector vector boson
(again, assumed to be the same for all gauge bosons for simplicity). One
can see that Yukawa couplings explicitly break $SU(2)_R$ in
composite sector (see Eq. (\ref{complagr})). But this breaking gives
a small contribution to the $T$ parameter and is thus technically natural as
mentioned in \cite{Contino:2006nn}.\footnote{Alternatively, we can
add extra composite site fermions so that Yukawa interactions respect
$SU(2)_R$. This corresponds to choosing
$5D$ fermions in complete multiplets of
$SU(2)_R$ in the $5D$
AdS
models \cite{Agashe:2003zs}. We will not pursue this
option here.}

\subsection{Mixing and Diagonalization}
The two sectors (composite and elementary) are connected to each other by the mixing terms
\begin{equation} \label{soft mix}
{\cal L}_{\small{\text{mixing}}} =
 - M_*^2\, \frac{g_{el}}{g_*}\, A_{\mu} \rho_{\mu}^* +
   \frac{M_*^2}{2} \left( \frac{g_{el}}{g_*} A_{\mu} \right)^2
   + (\bar\psi_L \Delta \chi_R + \bar{\tilde\psi}_R \tilde\Delta \tilde\chi_{L} + {\rm h.c.}).
\end{equation}
Due to the presence of the gauge boson mixing terms the following 
combination of the vector bosons will remain massless
\begin{equation}\label{SM gauge boson}
\frac{g_*}{\sqrt{g_{el}^2 + g_*^2}}\, A_{\mu} +
\frac{g_{el}}{\sqrt{g_{el}^2 + g_*^2}}\, \rho^*_{\mu}.
\end{equation}
The original elementary and composite states will be re-written using 
the mass eigenstates as follows
\begin{align}
\label{gauge mass diag}
\begin{pmatrix} A_\mu \\ \rho^*_\mu \end{pmatrix}  &\rightarrow
 \begin{pmatrix} \cos\theta & -\sin\theta \\ \sin\theta & \cos\theta \end{pmatrix}
 \begin{pmatrix} A_\mu \\ \rho^*_\mu \end{pmatrix}\, , \quad &
 \tan\theta &= \frac{g_{el}}{g_{*}}\, , \\[0.3cm]
\label{L fermion mass diag}
\begin{pmatrix} \psi_L \\ \chi_L \end{pmatrix}  &\rightarrow
 \begin{pmatrix} \cos\varphi_{\psi_L} & -\sin\varphi_{\psi_L} \\
   \sin\varphi_{\psi_L} & \cos\varphi_{\psi_L} \end{pmatrix}
 \begin{pmatrix} \psi_L \\ \chi_L \end{pmatrix}\, , \quad &
 \tan\varphi _{\psi_L} &= \frac{\Delta}{m_*}\, , \\[0.3cm]
\label{R fermion mass diag}
\begin{pmatrix} \tilde\psi_{R} \\ \tilde\chi_{R} \end{pmatrix}  &\rightarrow
 \begin{pmatrix} \cos\varphi_{\tilde\psi_{R}} & -\sin\varphi_{\tilde\psi_{R}} \\
   \sin\varphi_{\tilde\psi_{R}} & \cos\varphi_{\tilde\psi_{R}} \end{pmatrix}
 \begin{pmatrix} \tilde\psi_{R} \\ \tilde\chi_{R} \end{pmatrix}\, , \quad &
 \tan\varphi_{\tilde\psi_{R}} &= \frac{\tilde\Delta}{\tilde m_*}\, .
\end{align}
In the new, i.e., mass eigenstate basis, 
$(A_\mu,\psi_L,\tilde{\psi}_R)$ are the SM fields, which are
massless {\em before} EWSB, and $(\rho_*^\mu,\chi_L,\tilde{\chi}_R)$ are
the 
heavy mass eigenstates (i.e.
the heavy partners of SM), again prior to EWSB.
To shorten our notations we will denote
\begin{eqnarray}
&\theta\equiv\theta_1,\theta_2,\theta_3,~~\varphi_{\psi_L}\equiv\varphi_{q_{Li}},\varphi_{l_{Li}}
,~~\varphi_{\tilde\psi_R}\equiv\varphi_{u_{Ri}},\varphi_{d_{Ri}},\varphi_{\nu_{Ri}},\varphi_{e_{Ri}}\nonumber\\
&\sin \varphi_{u^i_R}\equiv s_u,~~
\sin\varphi_{d^i_R}\equiv s_d,~~
\sin \varphi_{q_{Li}}\equiv s_q.
\end{eqnarray}

\subsection{Couplings in mass eigenstates before EWSB}
Substituting Eq. (\ref{gauge mass diag}) (\ref{L fermion mass
diag}) (\ref{R fermion mass diag}) in Eq. (\ref{complagr}), we get
the Lagrangian for the Yukawa interaction between  quarks and
Higgs field in mass eigenstates before EWSB (the same expression
will be true for leptons too, one just has to substitute $L,E,N
\Longleftrightarrow Q,D,U $)

\begin{eqnarray}
\label{yukawa} {\cal L}_Y=&&{\cal L}^{\text{SM-SM}}_Y+{\cal
L}^{\text{SM-Heavy}}_Y+{\cal L}^{\text{Heavy-Heavy}}_Y\nonumber\\
=
&&-Y_{*u}\tilde{H} s_qs_u\bar{q}_L u_R-Y_{*d}{H} s_qs_d\bar{q}_L d_R\nonumber\\
&&-Y_{*u} \tilde{H} \left[ c_q s_u\bar{Q}_L u_R +s_q c_u \bar{q}_L \tilde{U}_R \right]-Y_{*d}H\left[c_q s_d\bar{Q}_L d_R +s_q c_d \bar{q}_L \tilde{D}_R\right]\nonumber\\
&&-Y_{*u}\tilde{H}\left[ c_qc_u\bar{Q}_L \tilde{U}_R+\bar{Q}_R \tilde{U}_L \right]-Y_{*d}{H}\left[ c_qc_d\bar{Q}_L \tilde{D_R} +\bar{Q}_R \tilde{D}_L \right]+\text{h.c.}
\end{eqnarray}
where $c_{q,u,d}$ stands for $\cos(\varphi_{q,u,d})$. We have
split the Yukawa interactions into three parts, (SM-SM):
interaction between two SM fermions, (SM-Heavy): interaction
between SM fermion and heavy fermions, and  (Heavy-Heavy):
interaction between two heavy fermions.

Similarly interactions between fermions (including SM and heavy)
and heavy partners
of SM gauge bosons are
\begin{eqnarray}
\label{gauge}
&& {\cal L}={\cal L}^{\text{SM-SM}}+{\cal L}^{\text{SM-Heavy}}+{\cal L}^{\text{Heavy-Heavy}}\nonumber\\
&& = \rho^*_\mu g\left[\bar q_L\gamma_\mu q_L(-c_q^2t+s_q^2\frac{1}{t})\right] \nonumber\\
&&+\rho^*_\mu g\left[(\bar q_L\gamma_\mu Q_L+\bar Q_L\gamma_\mu
q_L)( s_q c_q(1+\frac{1}{t}))\right]\nonumber\\
&& +\rho^*_\mu g\left[\bar Q_L\gamma_\mu Q_L(c_q^2 \frac{1}{t}
-s_q^2 t) \right]\nonumber\\
&&+\{ L\leftrightarrow R \},
\end{eqnarray}
where $t \equiv \tan\theta$, and $g$ is usual SM gauge coupling
constant, and it is equal to $g=g_{el} \text{cos} \theta=g_{*}
\text{sin}\theta$. In the same way as we have done for the Yukawa
interactions we split total Lagrangian into three parts ((SM-SM),
(SM-Heavy), (Heavy-Heavy)) . In the limit when all the SM fermions
are made up of mostly elementary sector particles, 
i.e. $s_q \ll 1$, then the flavor
non-universal interaction between SM quarks and heavy gauge bosons
will be $\frac{g
s_q^2}{\text{tan}\theta}=g_{*}s_q^2\text{cos}\theta\approx
g_{*}s_q^2$, and similarly for the right handed quarks.

The interactions between Higgs field, massless vector bosons and
their heavy partners are
\begin{eqnarray}\label{Higgsgauge}
{\cal L}=&&{\cal L}^{\text{SM-SM}}+{\cal
L}^{\text{SM-Heavy}}+{\cal L}^{\text{Heavy-Heavy}}=
|D_\mu H|^2\nonumber\\
&&+\left[H^{\dagger}ig \cot \theta \rho^*_\mu D_\mu H
-i\frac{g_1}{2\sin \theta_1 }\left(\frac{1}{\sqrt{2}}\tilde{H}^\dagger \tilde W_\mu^- D_{\mu} H+
\frac{1}{\sqrt{2}}{H}^\dagger \tilde W_\mu^+ D_{\mu} \tilde H -
\sqrt{\frac{3}{5}} H^\dagger \tilde{\cal B} D_\mu H \right )\right]\nonumber\\
&&+\left [
-g_1g\frac{\cot \theta}{2\sin \theta_1 } \left(\frac{1}{\sqrt{2}} \tilde{H}^\dagger \tilde W_\mu^- \rho^*_{\mu} H+
\frac{1}{\sqrt{2}}{H}^\dagger \tilde W_\mu^+
\rho^*_\mu \tilde H -
\sqrt{\frac{3}{5}} H^\dagger \tilde{\cal B} \rho^*_\mu H \right )\right.\nonumber\\
&&+\left. H^\dagger\left((g \cot\theta \rho^*_\mu)^2+
\frac{g_1^2}{\sin^2\theta_1}(\frac{1}{2}\tilde W_\mu^+\tilde W_\mu^-+\frac{3}{20}{\cal B}_\mu^2)\right)H \right]
\end{eqnarray}

\subsection{Flavor Anarchy}

We make the assumption that composite site Yukawa couplings are
``anarchical'', i.e., there is no large hierarchy between elements
within each matrix $Y_{\ast\,u,d}$. However, we need hierarchies in the
elementary/composite mixing angles ($s_{q,u,d}$) to
reproduce the hierarchical quark masses and CKM mixing angles. Such
a choice appears arbitrary from the point of view
of the two-site model, i.e., why some couplings are hierarchical and
others are not, but
%
%
this choice will be justified by the correspondence with the
5D model (see Appendix \ref{relationto5d}).

\subsection{Including EWSB}\label{EWSB}
Plugging in the Higgs vev in Eq. (\ref{yukawa})(\ref{Higgsgauge})
will lead to new mixings between SM massless fields and their heavy
partners which can be classified in the same way as was done in  Eq.
{\eqref{yukawa}},\eqref{gauge},\eqref{Higgsgauge}: (SM-SM)- mixing
between different generations of the SM massless fermions and the
mixing between  ($W^3,B$) SM  gauge fields ; (SM-Heavy)- mixing
between SM massless fermions and heavy fermions and the mixing
between  ($B,W^3$) SM gauge bosons and ($W^3_*,{\cal
B}_*,\tilde{{\cal B}_*},W_*$) heavy vector bosons; (Heavy-Heavy)-
mixing between the heavy fermions corresponding  to the different
generations of SM and the mixing between $(W^3_*,{\cal
B}_*,\tilde{{\cal B}_*},W_*)$ heavy vector bosons. These mixings
lead to many new contributions to flavor violating processes, which
we will study in detail in later sections.

\section{$\Delta F =2$ processes: $\epsilon_K$}\label{ek}
\subsection{Formulae for Two-Site Model}
\begin{figure}
\begin{center}
\includegraphics[scale=1]{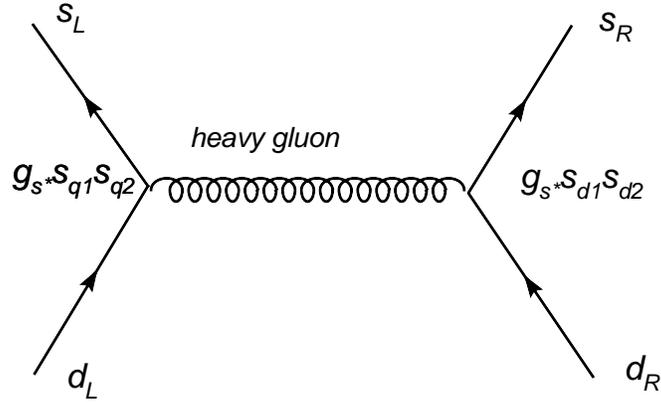}
\caption{\label{fig_gluon}  Feynman diagram for $\Delta S = 2$
process via heavy gluon exchange}
\end{center}
\end{figure}

We want to find the bound on composite sector scale from CP violation
in the $\Delta
S=2$ process, i.e., $\epsilon_K$. The most general effective
Hamiltonian for $\Delta S=2$ processes can be parameterized in the
following way \cite{Buras:1998raa}

\begin{eqnarray}
\label{hamiltonian}
&H_{ \Delta S = 2 } = C_1 {\cal O}_1+C_2 {\cal O}_2+ C_3 {\cal O}_3+ C_4 {\cal O}_4 + C_5 {\cal O}_5 \; \hbox{with} \nonumber\\
& {\cal O}_1=\bar{d}_{L}^\alpha\gamma_\mu
s_{L}^{\alpha}\bar{d}_{L}^\beta\gamma_\mu s_{L}^{\beta},~~
 {\cal O}_2=\bar{d}_{R}^\alpha s_{L}^{\alpha}\bar{d}_{R}^\beta s_{L}^{\beta}\nonumber\\
& {\cal O}_3=\bar{d}_{R}^\alpha s_{L}^{\beta}\bar{d}_{R}^\beta s_{L}^{\alpha},~~
 {\cal O}_4=\bar{d}_{R}^\alpha s_{L}^{\alpha}\bar{d}_{L}^\beta s_{R}^{\beta},~~
 {\cal O}_5=\bar{d}_{R}^\alpha s_{L}^{\beta}\bar{d}_{L}^\beta s_{R}^{\alpha},
\end{eqnarray}
where $\alpha, \beta$ are color indices. There are also $\cal
O^{ \prime }_{ 1, \; 2 }$ operators with $L$ replaced by $R$.
The dominant contributions to these Wilson coefficients in the
two-site model come from tree-level exchange of heavy gauge bosons
-- for example gluon (see Fig.
\ref{fig_gluon}) --
with flavor violating couplings.
These flavor violating couplings arise mainly from the mixings
between SM fermions induced after EWSB (see section \ref{EWSB})
which we now focus on -- the other two types of mixings (SM-Heavy,
Heavy-Heavy) have sub-leading effects for $\epsilon_K$
and so will be neglected for
the analysis in this section.
%
%

%
The point is that the couplings between heavy gluon and SM quarks
are diagonal but non-universal in the gauge eigenstate basis for
quarks, i.e., before EWSB, in ${\cal L}^{\text{SM-SM}}$ term of Eq.
(\ref{gauge}). After EWSB, one has to use unitary transformations:
($D_L$, $D_R$) and ($U_L$, $U_R$) to go to mass eigenstate basis for
down and up-type quarks respectively (just like in the SM). These rotations
thus lead to off-diagonal couplings between SM quarks (in mass
eigenstate basis) and heavy gluon. From the analysis of the $5D$
models \cite{Fitzpatrick:2007sa, Davidson:2007si, Csaki:2008zd,
Blanke:2008zb}, it is well-known that the dominant contribution
comes from the heavy/KK gluon exchange between left-handed and
right-handed down-type quark currents, i.e., $(V-A) \times(V+A)$-type
operators. Therefore, we focus here on heavy gluon exchange of the
above type. It is straightforward to show that such
exchange
gives (upon Fierzing)
%
%
\begin{eqnarray}\label{C4K}
C_4 \left( M_{\ast} \right) & = & -3 C_5 \left( M_{ \ast} \right) \nonumber \\
 & = & \frac{ (g_{s \ast })^2 }{ M_{ \ast }^2 }
\big[ ( s_{q2})^2
\left( D_L \right)_{ 1 2 } +
\left( s_{q3} \right)^2
\left( D_L \right)_{ 1 3 } \left( D_L \right)_{ 2 3 } \Big] \times \nonumber \\
& & \Big[
\left( s_{d2}\right)^2
\left( D_R \right)_{ 1 2 } +
\left( s_{d3}\right)^2
\left( D_R \right)_{ 1 3 } \left( D_R \right)_{ 2 3 } \Big]^*
\end{eqnarray}
where $g_{s \ast}$ is composite QCD coupling. Each $\Big[...\Big]$
in this formula includes
two terms, i.e., one from the ``direct'' $1-2$ mixing (present even with two
generations) and another from the
$(1-3) \times (2-3)$ mixing (i.e., via 3rd generation)
for the left and right-handed flavor-violating couplings.

Assumption of anarchic Yukawa couplings $Y_*$  in the original
Lagrangian of Eq. (\ref{yukawa}) implies that mixing angles in SM
Yukawa couplings are given by ratios of elementary-composite
mixings\cite{Agashe:2004cp}, for example,
\begin{eqnarray}
\label{mixing angles} \left( D_{L,R} \right)_{ i j } & \sim & \frac{
( s_{q,d})_i }{ (s_{q,d})_j} \; \hbox{for} \; i < j
\end{eqnarray}
So, the two terms (inside each of the brackets $\Big[...\Big]$) 
in Eq. (\ref{C4K})
(for each of left and right-handed sectors) are of same size, but
uncorrelated.

On the other hand from the ${\cal L}^{\text{SM-SM}}_Y$ term of  Eq. (\ref{yukawa})
we have
\begin{eqnarray}\label{quarkmass}
m_d\sim Y_*^d s_{q1}s_{d1} v/\sqrt{2}\\ \nonumber m_s\sim Y^d_*
s_{q2}s_{d2} v/\sqrt{2},
\end{eqnarray}
so we can estimate the size of the mixing angles $s_{qi},s_{di}$.

Now we can estimate new physics contribution to $C_{ 4, \; 5 }$ using the
following assumptions: (i) considering one term in each of the
brackets $\Big[...\Big]$
of Eq. (\ref{C4K}) at a time, (ii) mixing angles set to ``natural'' size
(i.e., with ``$=$'' in Eq.(\ref{mixing angles}) above), and (iii)
quark masses given by natural size of the parameters
(i.e., with ``$=$'' in
Eq.(\ref{quarkmass}) above). Plugging Eq. (\ref{mixing angles}) and
(\ref{quarkmass}) into Eq. (\ref{C4K}) leads to the estimate, up to an
$O(1)$ complex factor:
\begin{eqnarray}
C^{ 2-\text{site} }_{ 4 \; \hbox{ estimate} } & = & \frac{ g_{ s
\ast }^2 }{ (Y^d_{ \ast })^2 } \frac{ 2 m_s m_d }{ v^2 } \frac{1}{
M_{ \ast }^2 } \label{modeleps}
\end{eqnarray}
with $v = 246$ GeV, where subscript ``estimate'' stands for the
above three  assumptions. To repeat, the assumption of anarchy tells us that the
four terms in Eq. (\ref{C4K}) are of the same size
as Eq. (\ref{modeleps}) and have
{\em un}correlated phases. Therefore, our estimate using one term gives us
the correct result up to $O(1)$ factor.

\subsection{Experimental limit}

The model independent bound from $\epsilon_K$ is strongest on the
Wilson coefficient $C_{4}$ due to (i) enhancement
(as compared to for
the other Wilson coefficients) from
RG scaling from the new
physics scale to the hadronic scale and (ii) from chiral enhancement
of matrix element (see reference \cite{Bona:2007vi}).\footnote{The effect of
$C_5 \left( M_{ \ast } \right)$ in the two-site model is
sub-leading because firstly the
model-independent bound is weaker relative to $C_4$
(see reference \cite{Bona:2007vi}) and
secondly in this model $C_5 \left( M_{ \ast } \right)$ is suppressed by a color factor relative to $C_4$ (see Eq.
\ref{C4K}).} This bound on $C_4$ is :
\begin{eqnarray}
\hbox{Im}\,C_4 & \stackrel{<}{\sim} & \frac{1}{ \left( \Lambda_F
\right)^2}, \qquad \Lambda_F =  1.6 \times 10^5  \,\hbox{TeV}.
\label{expteps}
\end{eqnarray}
where the coefficient is renormalized at the $\sim 3$ TeV
scale \cite{Csaki:2008zd}. 
Note that
the bound on Im $C_4$ is only mildly (logarithmically) sensitive to
the renormalization scale and hence it remains almost the same 
as the above number (which is again for a scale of $\sim 3$ TeV)
for
heavy mass scales of up to $\sim 10$ TeV that we will consider in this paper.
Using Eqs. (\ref{modeleps}) and (\ref{expteps}), and assuming order
one phase,
we get
\begin{eqnarray}
\label{epsilonk} M_{ \ast } \stackrel{>}{\sim} \frac{11 g_{ s\ast}
}{ Y^d_{ \ast }} \; \hbox{TeV}
\end{eqnarray}
We can see the bound on the composite mass scale {\em decreases} as $Y_*^d$
increases.

\section{Radiative processes: $ b \rightarrow s \gamma$}\label{bsg}

The rare decay $B\rightarrow X_s \gamma$ gives very powerful
constraints on new physics. We follow the standard notation and
define the effective Hamiltonian for $b \rightarrow s \gamma$
\cite{Buras:1998raa}:
\begin{equation}
\mathcal{H}_{eff}(b\rightarrow s \gamma) = -
\frac{G_F}{\sqrt{2}}V^*_{ts}V_{tb}[C_7(\mu_b) Q_7 +
C^{\prime}_7(\mu_b) Q^{\prime}_7+ \ldots]
\end{equation}
where $Q_7 = e \; m_b / \left( 8\pi^2 \right) \bar{b} \sigma^{\mu\nu} F_{\mu
\nu} (1-\gamma_5)s$ and $Q^{\prime}_7 = m_b \; e \left( 8\pi^2 \right) \bar{b}
\sigma^{\mu\nu} F_{\mu \nu} (1+\gamma_5)s$.
%
%
Here we have neglected other operators that only enter through
renormalization of $C_7$ and $C^{\prime}_7$. In SM, the Wilson
coefficient $C_7(\mu_w)$ evaluated at weak scale
is\cite{Buras:1998raa}
\begin{eqnarray}\label{SMbsg}
C^{ SM } _7(\mu_w) = -\frac{1}{2} \left[-\frac{(8 x_t^3 + 5 x_t^2 - 7
x_t)}{12(1-x_t)^3}+ \frac{x_t^2(2-3x_t)}{2(1-x_t)^4}\ln(x_t)
\right]; \qquad C^{\,\prime \; SM }_7(\mu_w) = \frac{m_s}{m_b}C^{ SM }_7(\mu_w)
\end{eqnarray}
with $x_t =  m_t^2/M_w^2$. The Wilson coefficient
$C^{\,\prime}_7(\mu_w)$ can be neglected in SM due to a suppression
by $m_s/m_b$. The leading order QCD correction gives
us \cite{Buras:1998raa}
\begin{eqnarray}\label{c7}
C_7(\mu_b)= 0.695 C_7(\mu_w) + 0.085 C_8(\mu_w)-0.158 C_2(\mu_w)\\
\nonumber = 0.695(-0.193)+0.085(-0.096) -0.158=-0.300
\end{eqnarray}
where $C_2$ and $C_8$ are Wilson coefficients for operators $Q_2
\equiv (\bar{c}b)_{V-A}(\bar{s}c)_{V-A}$ and $Q_{8G} \equiv \\
m_b \; g / \left( 8 \pi^2 \right) \bar{b}_\alpha \sigma^{\mu \nu} (1-\gamma_5)
T^a_{\alpha \beta}s_\beta G^a_{\mu\nu}$.
The latest higher order calculations for BR($b\longrightarrow s
\gamma$) are given in \cite{Misiak:2006zs} but the above order results
suffice for our purposes.

\subsection{Estimate in two-site model}
%
%
\label{bsgonegen}

\begin{figure}
\begin{center}
\includegraphics[]{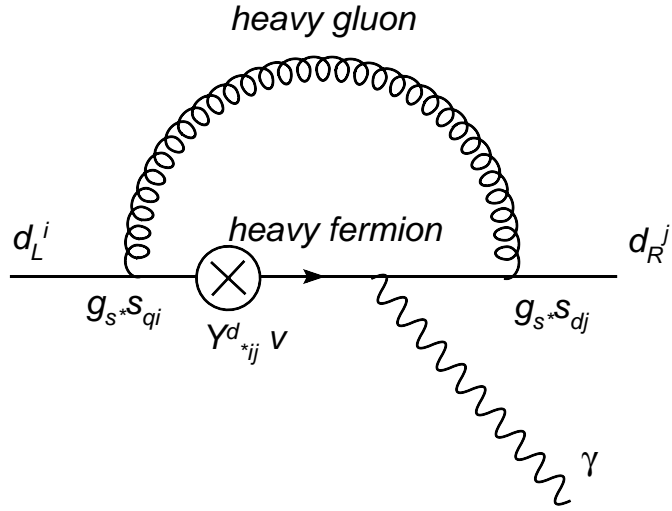}
\caption{\label{kkgluondiag}  Feynman diagrams for b $\rightarrow$ s
$\gamma$ via heavy gluon and heavy fermions}
\end{center}
\end{figure}

In two-site model, the largest new physics contribution to
$\Gamma(b \rightarrow s \gamma)$ comes from diagrams with heavy
states in the loop because of their larger coupling constants.
First, we consider diagrams with heavy gluons and fermions (see
Fig. \ref{kkgluondiag}). We can get an idea of the flavor
structure of this diagram by treating the
EWSB-induced fermion mass terms of Eq. (\ref{yukawa}) as being small
compared to the masses of
the heavy partners of SM fermions (henceforth called by
the mass insertion approximation).
From ${\cal{L}}^{\text{SM-Heavy}}$ term of Eq. (\ref{gauge}), we
see that mass insertion approximation gives us a new contribution
to Wilson coefficients of operators $\bar{d_j} \sigma^{\mu\nu}
F_{\mu \nu} (1-\gamma_5)d_{i}$ (with quarks in gauge basis before
EWSB)
\begin{eqnarray}
C^{G}_{7\, ij} \propto s_{q_i} g_{s\ast}^2 Y^d_{\ast ij} s_{d_j}
\end{eqnarray}
Notice that $C^G_{7\, ij}$ has the same flavor structure as quark
mass matrix $m_{d\, ij} \approx Y^d_{\ast\, ij} s_{q_i} s_{d_j}$.
Therefore, after unitary rotation into the mass eigenstates after EWSB,
$C^G_{7\, ij}$ will be approximately diagonal in flavor space, 
and contribution from
heavy gluon and heavy fermion exchange to $\Gamma(b\rightarrow s
\gamma)$ is suppressed. (see reference \cite{Agashe:2004cp} for a similar
discussion in warped extra dimension, where KK gluons and KK
fermions correspond to heavy gluons and fermions here.)

\begin{figure}
\begin{center}
\includegraphics[]{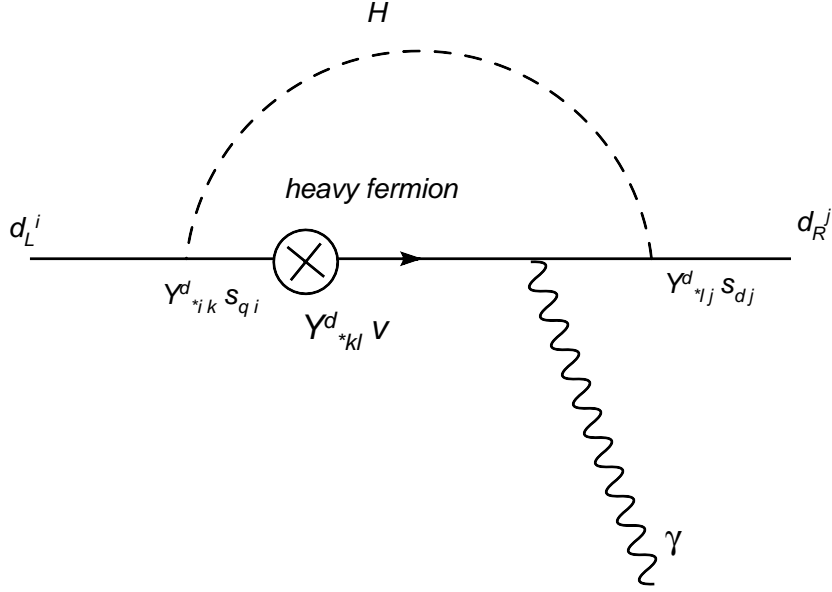}
\caption{\label{figinsertion} Feynman diagrams for b $\rightarrow$ s
$\gamma$ via Higgs using mass insertion}
\end{center}
\end{figure}

Next, we consider diagrams with heavy fermions and Higgs in the loop
(including physical Higgs and longitudinal W/Z bosons). Similar to
the previous analysis, we can get the flavor structure of these
diagrams from mass insertion approximation. For the purpose of
estimating flavor structure, we consider only neutral Higgs diagram
(see Fig. \ref{figinsertion}). From the Yukawa couplings between SM
fermion, heavy fermion and Higgs (${\cal{L}}^{\text{SM-Heavy}}$ term
of Eq. \ref{yukawa}), we find that
\begin{eqnarray}
C^H_{7\, ij} \propto s_{q_i} Y^d_{\ast ik} Y^d_{\ast kl} Y^d_{\ast
lj} s_{d_j}
\label{flavorstrucH}
\end{eqnarray}
It is obvious that $C^H_{7\, ij}$ is not aligned with $m_{d\, ij}$,
assuming no particular structure in the $Y_{ \ast }$ (i.e.,
anarchy).
Thus these diagrams will give the leading new contribution to $C_7$ and
$C^\prime_7$, and we will focus on these diagrams (see reference
\cite{Agashe:2004cp} for a similar discussion in warped extra
dimension).

Because of the near degeneracy of heavy fermion masses, we can{\em not}
use mass insertion approximation to {\em calculate} the loop diagrams.
Instead, we need to diagonalize the $9\times 9$ mass matrix (once 
we include EWSB-induced mass terms, i.e., 
coming
from Yukawa couplings in Eq. (\ref{yukawa})) for all down type
quarks
in order to determine the mass eigenstates and their
couplings.
Since it
is difficult to obtain an exact analytical formulae
for this effect, the
analysis is performed numerically in Section \ref{numerical}.
However, it is insightful to obtain an approximate
analytical formulae for $b \rightarrow
s \gamma$ as follows.
First, we
calculate the dipole operator for the case of {\em one} generation quark
together
with its heavy partners (say, as in the calculation of $(g-2)_{\mu})$) 
with{\em out} using the mass insertion approximation
and
then we simply multiply it by
%
%
factors from generational mixing
effects in order to obtain the amplitude for
$b \rightarrow s \gamma$.

In more detail, we diagonalize the $3\times 3$ mass matrix 
(including the EWSB-induced mass terms) for one
generation quarks analytically to first order in $x \equiv
Y^{u,d}_\ast v / \left(
m_*\sqrt{2} \right)$ in Appendix \ref{B}: the results for dipole
moment operator of one generation with charged and neutral Higgs in
the loop are shown in Eqs. (\ref{Eq.charged Higgs}) and
(\ref{Eq.neutral Higgs}). In order to {\em estimate} the effect of mixing
between different generations, we again use mass insertion
approximation
%
%
(see Fig. \ref{figinsertion}). For
example, the operator $\bar{b}_L \sigma^{\mu\nu}F_{\mu\nu} s_R$ can
be generated via the mass insertions/Yukawa couplings
(as in Eq. \ref{flavorstrucH},
but dropping the flavor indices on $Y_{d*}$
for simplicity)
\begin{equation}\label{Eq. mixing1}
Y_{d*} s_{q3} Y_{d*} v Y_{d*} s_{d2}
\end{equation}
Based on our assumption of anarchy and the formulae for Yukawa couplings
and mixing angles (Eq. (\ref{yukawa}) (\ref{mixing angles})), we
know that
\begin{equation}
Y_{d*} v s_{q3} s_{d2} = Y_{d*} v s_{q3} s_{d3}
\frac{s_{d2}}{s_{d3}} \sim m_b (D_R)_{23}
\end{equation}
and
\begin{eqnarray}
\frac{ m_s }{ m_b } & \sim & \left( D_L \right)_{ 23 } \left( D_R
\right)_{ 23 }
\end{eqnarray}
In addition, since left-handed down and up-type quarks have the same
elementary-composite mixing, we get (again assuming anarchy of $Y_{ d\ast
} $)
\begin{eqnarray}\label{Eq.mixing2}
\left( D_L \right)_{ 23 } & \sim & \left( U_L \right)_{ 23 } \nonumber \\
 & \sim & V_{ ts } \;
\hbox{or} \; V_{ cb }
\end{eqnarray}
where in the second line we have used that $V_{ CKM } = U_L^{
\dagger } D_L$. Combining Eq (\ref{Eq. mixing1}) through
(\ref{Eq.mixing2}), we can find that generational mixing gives a
factor $\sim \frac{m_s}{m_b V_{ts}}$. Similarly, for the operator
$\bar{b}_R \sigma^{\mu\nu}F_{\mu\nu} s_L$ we have (as in Eq. \ref{flavorstrucH})
\begin{equation}
Y_{d*} s_{d3} Y_{d*} v Y_{d*} s_{q2} \sim (Y_{d*})^2 m_b (D_L)_{23}
\sim (Y_{d*})^2 m_b V_{ts}
\end{equation}
i.e., generational mixing gives a factor $\sim V_{ts}$. Note that
for neutral Higgs diagram the amplitude is proportional to
$Y_{d\ast}^3$. The flavor structure for {\em charged} Higgs (would-be
Goldstone) diagram is similar, expect that there are two types of
contributions (schematically
$\propto Y_{d\ast}^3$ and $Y_{u\ast}^2 Y_{d\ast}$). For
simplicity, we set $Y_{u\ast} = Y_{d\ast} \equiv Y_*$ in our
estimation.

Then, multiplying the one generation results for dipole
operator in Eqs. (\ref{Eq.charged Higgs}) and (\ref{Eq.neutral
Higgs}) by the above generational mixing factors, we get the following
effective Hamiltonians:
\begin{equation}\label{Eq.charge eff operator}
\mathcal{H}^{eff}_{\text{charged\, Higgs}} \approx
\frac{5}{12}(Y_*)^2m_b\frac{i e}{16 \pi^2} \frac{(2\epsilon\cdot
p)}{(m_*)^2} [ V_{ts}\bar{b} (1-\gamma_5) s+ \frac{m_s}{m_b V_{ts}}
\bar{b} (1+\gamma_5) s]
\end{equation}
\begin{equation}
\mathcal{H}^{eff}_{\text{neutral\, Higgs}} \approx
-\frac{1}{4}(Y_*)^2 m_b\frac{i e}{16 \pi^2} \frac{(2\epsilon\cdot
p)}{(m_*)^2} [ V_{ts} \bar{b} (1-\gamma_5) s+  \frac{m_s}{m_b
V_{ts}}\bar{b}(1+\gamma_5) s]
\end{equation}
We present the results for both charged Higgs and neutral Higgs
contribution since they generally have different phase and cannot be
simply added together. Since their sizes are of the same order, we
will {\em focus just on charged Higgs contribution in the analytical
estimates}. Then, the new physics contribution to the Wilson coefficients
are\footnote{Note that such a size for these
Wilson coefficients can be {\em estimated}, i.e., derived
up to $O(1)$
factors, using {\em purely} 
mass insertion approximation. As explained above, here
instead we have {\em calculated} the $O(1)$ factor from loop
diagram
(with{\em out} using mass insertion approximation), although
we still used mass insertion approximation to estimate the
generational mixing factors.}
\begin{eqnarray}
C^{2-\text{site}}_{7\; \hbox{estimate}}(m_*) =
-\frac{5}{48}\frac{(Y_*)^2}{(m_*)^2}\frac{\sqrt{2}}{G_F}; \qquad
C^{\,\prime\, 2-\text{site}}_{7\; \hbox{estimate}}(m_*) =
-\frac{5}{48}\frac{(Y_*)^2}{(m_*)^2}\frac{\sqrt{2}}{G_F}
\frac{m_s}{m_b \lambda^4} \label{estimatebsg}
\end{eqnarray}
where we used $V_{ts}\sim \lambda^2$ ($\lambda \approx 0.22$). As
explained earlier, (based on assumption of anarchy) in the exact
result for $b \rightarrow s \gamma$ there will be several terms of
the above order but with uncorrelated phases. Thus Eq.
(\ref{estimatebsg}) is only an {\em estimate} for $b \rightarrow s
\gamma$, i.e., the natural size of {\em one} term that contribute to
the new physics effective Hamiltonian. We expect the final result of
the coherent sum of such terms to be of the same order as this
one-term estimates. From these estimates we can conclude that
$C^{\,\prime 2-\text{site}}_7(m_*)$ is bigger than
$C^{2-\text{site}}_7(m_*)$ by a factor of $m_s / \left( m_b V_{ts}^2 \right)
\sim 8$, which is different than the case in SM (where
$C^{\,\prime}_7 \approx  C_7 \; m_s / m_b$).

As mentioned earlier, 
in Section \ref{numerical}, we will apply the exact diagonalization
of the $9 \times
9$ mass matrix for three generations to the results
from general loop calculation of
$b \rightarrow s \gamma$ in Appendix \ref{A} to obtain
$C_7^{2-\text{site}}$ and $C^{\,\prime 2-\text{site}}_7$
numerically.

\subsection{Experimental limit}

The leading order QCD corrections will suppress the new physics
contribution to the Wilson coefficients
\begin{eqnarray}
C^{2-\text{site}}_7(\mu_w) = \left[
\frac{\alpha_s(m_*)}{\alpha_s(m_t)} \right]^{16/21}\left[
\frac{\alpha_s(m_t)}{\alpha_s(\mu_w)} \right]^{16/23}
C_7^{2-\text{site}}(m_*) \approx 0.73\ C_7^{2-\text{site}}(m_*)
\end{eqnarray}
We add it to $C^{ \text{SM} }_7(\mu_w)$ in Eq. (\ref{SMbsg}) and
then use this sum, i.e., $C^{ \text{total} }_7(\mu_w) = C^{
\hbox{SM} }_7(\mu_w) + C^{2-\text{site}}_7(\mu_w)$ in Eq. (\ref{c7})
to obtain $C_7(\mu_b)$. Whereas, the SM contribution to $C^{\,
\prime}_7$ is negligible compared to that in the two-site model so
that we have
\begin{eqnarray}
C^{\,\prime \text{total} }_7(\mu_b) & \approx &
C^{\,\prime 2-\text{site} }_7(\mu_b) \nonumber \\
& = &
\left[
\frac{\alpha_s(m_*)}{\alpha_s(m_t)} \right]^{16/21}\left[
\frac{\alpha_s(m_t)}{\alpha_s(\mu_b)} \right]^{16/23} C^{\,\prime
2-\text{site}}_7(m_*) \nonumber \\
& \approx & 0.48\  C^{\,\prime
2-\text{site}}_7(m_*)
\end{eqnarray}
The contributions from $C_7(\mu_b)$ and $C^{\,\prime}_7(\mu_b)$ sum
incoherently (without interference) in the total (i.e., SM
and new physics) decay width
$\Gamma^{\text{total}}(b \rightarrow s \gamma)$:
\begin{equation}
\Gamma^{\text{total}}(b\rightarrow s\gamma) \propto |C_7(\mu_b)|^2 +
|C^{\,\prime}_7(\mu_b)|^2
\end{equation}
For convenience, we define $\delta_7 \equiv
C^{2-\text{site}}_7(m_*)/C^{SM}_7(\mu_w)$ and $\delta'_7 \equiv
C^{\,\prime\,2-\text{site}}_7(m_*)/C^{SM}_7(\mu_w)$. Adding these
new contributions, we have
\begin{equation}
\frac{\Gamma^{\text{total}}(b \rightarrow s \gamma)}{\Gamma^{SM}(b
\rightarrow s \gamma)} \approx 1 + 0.68 Re(\delta_7) +
0.11|\delta'_7|^2
\end{equation}
The experimental average value for the branching ratio is $BR(b
\rightarrow s \gamma) = (352 \pm 23 \pm 9) \times
10^{-6}$\cite{HFAG}. The theoretical calculation gives $BR(b
\rightarrow s \gamma) = (315 \pm 23) \times
10^{-6}$\cite{Huber:2007in}. Adding the $2\sigma$ uncertainties by
quadrature we find that a $20 \%$ deviation from SM prediction is
allowed.
If we consider the two contributions separately, we will get the
bound $|\delta'_7| \lesssim 1.4$ and Re$(\delta_7) \lesssim 0.3$.
Using Eqs. (\ref{estimatebsg}) and (\ref{SMbsg}), the first condition gives
\begin{equation}
\label{bsgamma} m_* \gtrsim (0.63) Y_*\, \hbox{TeV}
\end{equation}
and the second condition gives us a weaker bound. From this rough
estimate, we can see the bound on composite mass scale {\em increases}
with composite Yukawa coupling. \\

\subsection{Tension and
%
%
lowest heavy SM partner mass scale
scenario}\label{tension}

We see that the bounds on $M_*$ and $m_*$ from $\epsilon_K$ and
BR$\left( b \rightarrow s \gamma \right)$ have opposite dependence
on $Y_{\ast}$. Thus we cannot use this parameter to decouple
flavor-violation. {\em For simplicity, we set $M_* = m_*$ henceforth}. Then the
lowest allowed value for $M_*$ that satisfies both bounds
Eqs.(\ref{epsilonk}) and (\ref{bsgamma}) is
\begin{eqnarray}\label{bounds}
M_{ \ast } & \gtrsim & 2.6\sqrt{ g_{s \ast } } \; \hbox{TeV} \; \quad
\hbox{for} \quad Y_{ \ast } \sim 4.2 \sqrt{ g_{s \ast } }
\nonumber \\
& \sim & 4.5 \; \hbox{TeV} \; \hbox{for} \; g_{s \ast } \sim 3 \nonumber \\
& \sim & 6.4 \; \hbox{TeV} \; \hbox{for} \; g_{s \ast } \sim 6
\end{eqnarray}
where in last two lines, we have set $g_{s \ast } \sim 3, 6$ which
is motivated by the $5D$ AdS model,
although the latter value might not be allowed
by $5D$ perturbativity (see Appendix \ref{matching qcd}). We can
check that with the
%
%
values of $Y_*$ in Eq. (\ref{bounds}),
the loop expansion parameter  $Y_*^2 / \left( 16
\pi^2 \right)$ is less than one, and the
two-site model is thus perturbative (but barely so in the case of
$Y_{ \ast } \sim 10$ for
$g_{ s \ast } \sim 6$):
see
Appendix \ref{perturbativity} about perturbativity bound on KK
Yukawa couplings in the $5D$ AdS model.

We reiterate that the bounds
in Eq. (\ref{bounds}) are only
{\em estimates} in the sense that they are based on {\em one} among
multiple, {\em uncorrelated} terms in the amplitudes for both
$\epsilon_K$ and $b \rightarrow s \gamma$.
Also, note that the contributions
to $b \rightarrow s \gamma$ in the two-site model, being at the loop-level
(as opposed to the
tree-level contributions to $\epsilon_K$), can be
quite sensitive
to the composite sector content
-- for example, as mentioned in section \ref{review}, we could add
$SU(2)_R$ partners for the composite site $u_R$ and $d_R$ (as in $5D$ models)
which can easily modify the new physics amplitude for
$b \rightarrow s \gamma$
by $\sim O(1)$ factors due to their appearance in the loops. In this sense,
the constraints from $b \rightarrow
s \gamma$
presented for this model should especially be considered as
a ballpark guide to the viable parameter space of this framework:
the main motivation
for using $b \rightarrow s \gamma$ in our analysis
is to put an upper bound on the composite site Yukawa coupling.

As discussed in references \cite{ Agashe:2004cp,Agashe:2004ay} for
the $5D$ model, the Higgs-heavy fermion loop contributions to
electric dipole moments (EDMs) of SM fermions also increase with
the size of the composite Yukawa coupling (just like $b \rightarrow s
\gamma$). Thus, EDMs can also be used to put an upper bound on the
size of this coupling (for a given heavy mass scale). However, EDMs
depend on a different (flavor-{\em preserving}) combination of
phases than the flavor-violating observables $\epsilon_K$ and $b
\rightarrow s \gamma$ and so we will leave a study of these
constraints for the future. Note that $5D$ flavor symmetries can
suppress EDM's as well as the flavor violating effects.

\section{Correction to $Z b \bar{b}$ coupling}\label{Zbb}

There is another important constraint coming from non-universal
correction to $Z b_L \bar{b}_L$ coupling which arises from mixing
between SM and heavy states after EWSB (see \cite{Contino:2006nn})

\begin{equation}\label{Eqzbb}
\frac{\delta g_{Z\bar{b}b}}{g_{Z\bar{b}b}} \approx \sum_{i=1}^3
\left( \frac{Y_{\ast d i3}}{Y_{\ast u 33}}\right)^2 \left(
\frac{m_t}{M_\ast s_{u3}}\right)^2 + \frac{1}{2}\left(
\frac{m_t}{M_\ast s_{u3}}\right)^2 \left(\frac{g_{\ast 2}}{Y_{\ast
U33}} \right)^2
\end{equation}
Experimentally, it is measured to have less than $0.25\%$ deviation
from its SM value. If we assume that all composite Yukawa couplings
are of the same order, then we can get a bound on $M_\ast$ from the
first term alone:
\begin{equation}
M_\ast \gtrsim 4.7\quad \hbox{TeV}
\label{Zbbbound}
\end{equation}
This bound is similar to what we found from $\epsilon_K$ and $b
\rightarrow s \gamma$. However, if we allow a little hierarchy
between the Yukawa couplings, e.g., $Y_{\ast d} > Y_{\ast u}$,
then the bound on $M_\ast$ will be enhanced.
We mention that $Z\bar{b}_L b_L$ coupling can be protected by
another custodial symmetry\cite{Agashe:2006at}. But we will not use this
idea here.

\section{Numerical Analysis}\label{numerical}

In previous sections we presented semi-analytical {\em estimates} for
the new physics contributions to the $\epsilon_K$ and $b\rightarrow
s \gamma$ processes, but to get the precise values one has to perform
a numerical scan over the
parameter space.  The scan procedure is discussed in detail in
Appendix \ref{scan}. Here we summarize some
important features and results of our scan. We require that our composite
Yukawa coupling matrices are anarchical, i.e. all
entries of the same order,
with the results presented here corresponding to the variation of the Yukawa
couplings by a factor of three, and we varied the elementary/composite
mixings also by a factor of three.
First, we generate the points in parameter space with $Y_{u*},
Y_{d*}, s_Q, s_u, s_d$ such that the SM quark masses and CKM mixing
angles are reproduced. Then we calculated $|\frac{\Gamma^{\text
{total}}(b \rightarrow s \gamma)}{\Gamma^{\text {SM}}(b \rightarrow
s \gamma)}-1|/(20\%)$, $|\delta g_{Z\bar{b}b}/g_{Z\bar{b}b}|$ and
$\hbox{Im}\,C_{4K} \Lambda_F^2$ (with $\Lambda_F = 1.6 \times 10^5$
TeV) for different values of $M_*$ and $Y^{u,d}_*$.

\begin{figure}
\begin{center}
\includegraphics[scale = 0.7]{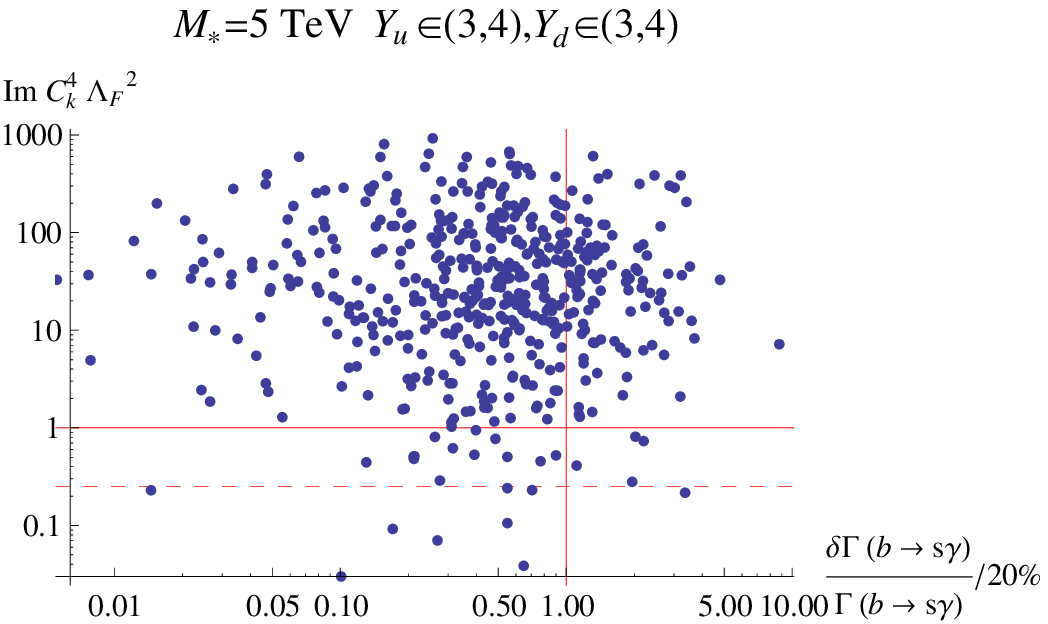}
\includegraphics[scale = 0.7]{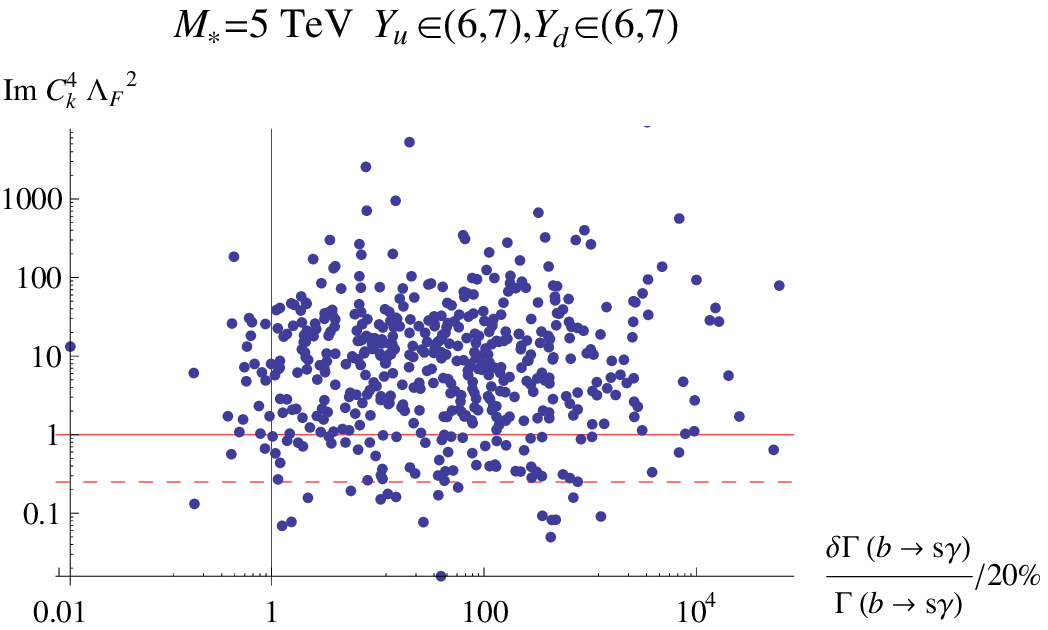}
\includegraphics[scale = 0.7]{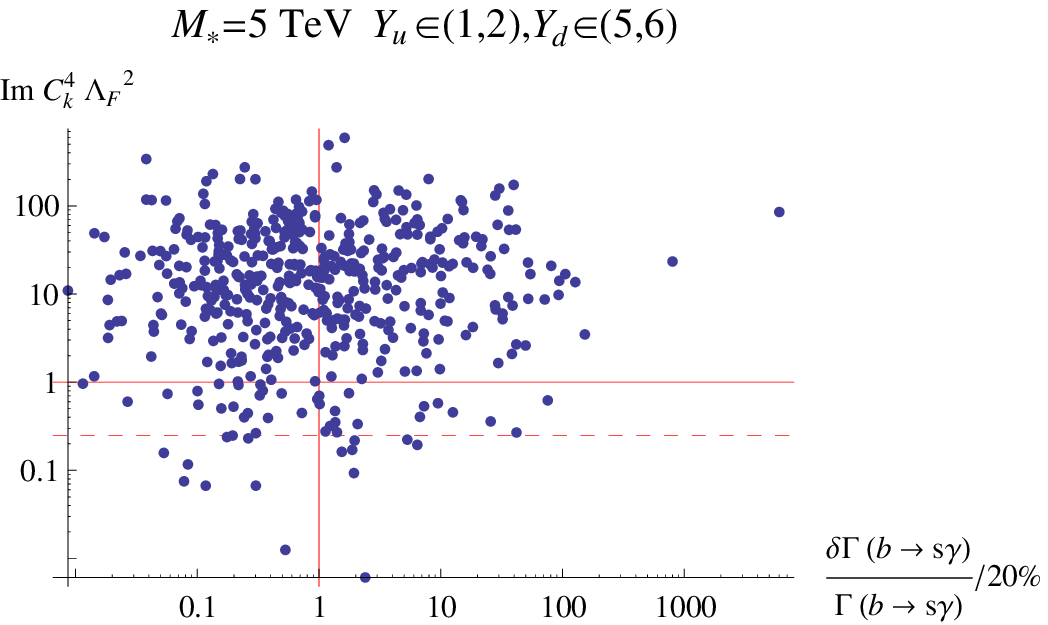}
\caption{\label{ekbsgm5} Scatter plot for shift in
$\hbox{BR}(b \rightarrow
s \gamma)$ and $\hbox{Im}\,(C_{4K})$ for $M_* = 5\ \hbox{TeV}$,
the
composite site gauge coupling
$g_{s*} = 3$ and different values of $Y_*^{u,d}$ (defined here as
the geometric {\em mean} of
the composite site
Yukawa couplings
$|Y_{*\,ij}^{u,d}|$). The allowed region is below and to the left
of the (red) solid lines.
For $g_{s*} = 6$, the allowed
region is below the
%
%
dashed line
and to the left of the solid (red) line.
(see discussion in section \ref{numerical}).}

\end{center}
\end{figure}

In Fig. $\ref{ekbsgm5}$, we show the plots of $|\frac{ 
\Gamma^{ \text{total} (b \rightarrow s \gamma) } }{ \Gamma^{\text SM}(b \rightarrow s
\gamma)}-1|/(20\%)$ and $\hbox{Im}\,C_{4K} \Lambda_F^2$ for $M_* =
5\  \hbox{TeV}$ and different values of ${Y}_{*}^{u,d}$ (defined here
as the geometric
{\em average} value for ${Y}_{*ij}^{u,d}$). We focus on the case with
$g_{s\ast} = 3$.
Points to the left and below the solid lines
satisfy both bounds from BR $\left( b \rightarrow s \gamma \right)$ and
$\epsilon_K$.
We begin with the cases with
no hierarchy between the up and down-type quark composite site
Yukawa coupling, i.e., ${Y}_\ast^d = {Y}_\ast^u$. In the top
left plot, we choose this value to be $\in (3,4)$. We see that a small
%
%
fraction of points
satisfy the bounds from $\epsilon_K$ and
BR $\left( b \rightarrow s \gamma \right)$.
Next we increase the common value for
${Y}_\ast^d$ and ${Y}_\ast^u$ to $(6,7)$ (top right plot). We expect
that the larger Yukawa coupling will enhance the contribution to
$\Gamma(b\rightarrow s\gamma)$ and suppress the contribution to
$\hbox{Im}\,C_{4K}$, which is clearly shown in the plots and illustrates
the tension discussed in section \ref{tension}. In the
end, there are fewer points satisfying both bounds with these larger
Yukawa couplings.

\begin{figure}
\begin{center}
\includegraphics[scale = 0.7]{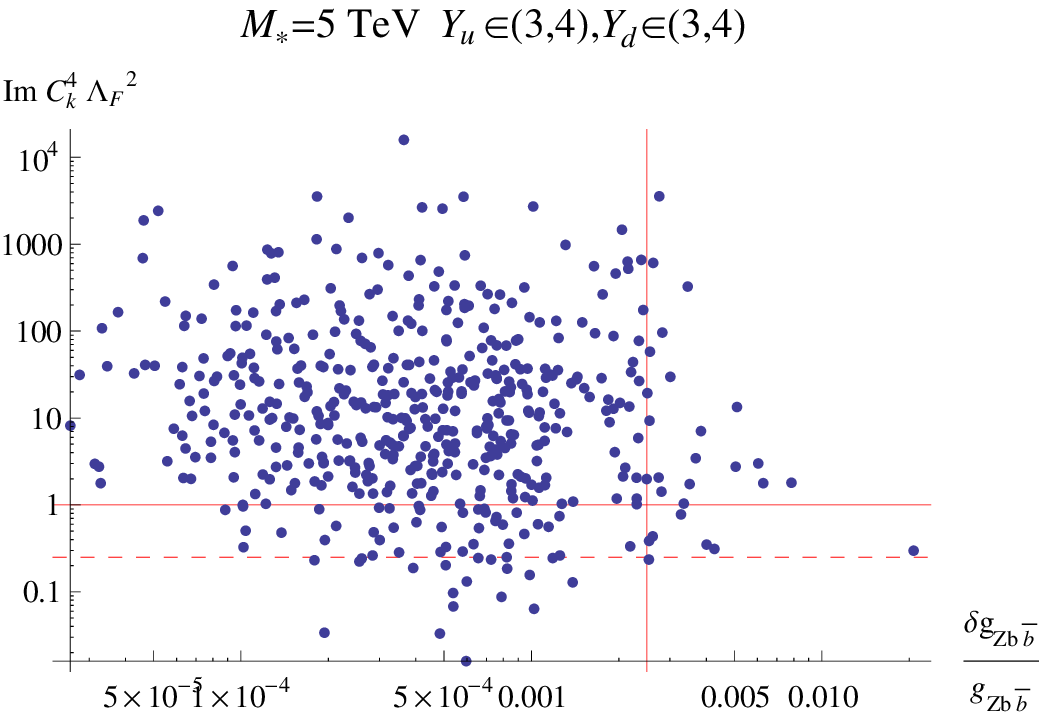}
\includegraphics[scale = 0.7]{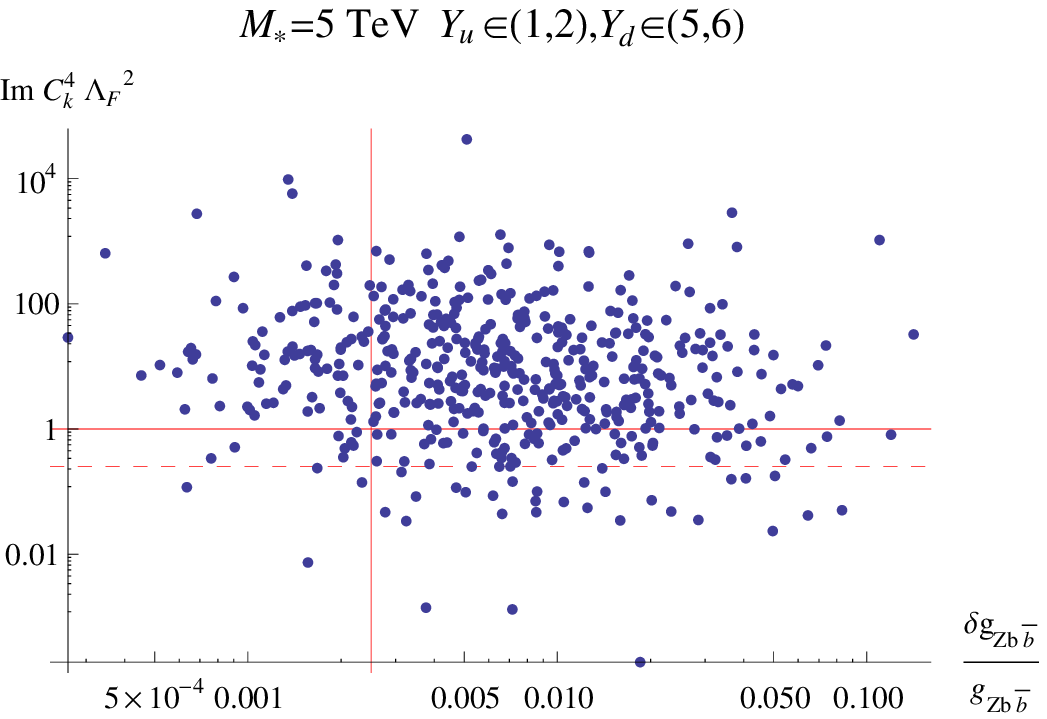}
\caption{\label{zbbek5}Scatter plot for $\delta g_{Z\bar{b}_L b_L}$
and $\hbox{Im}\,(C_{4K})$ for $M_* = 5\ \hbox{TeV}$, the
composite site gauge coupling $g_{s*} = 3$
and for different values of $Y_*^{u,d}$ (defined here as
the geometric {\em mean} of
the composite site
Yukawa couplings $|Y_{*\,ij}^{u,d}|$).
The allowed region is below and to the left of
the (red) solid lines. For $g_{s*} = 6$, the allowed
region is below the
%
%
dashed line
and to the left of the solid (red) line.
(see discussion in section \ref{numerical}).}
\end{center}
\end{figure}

Finally, we consider a mild hierarchy between the Yukawa
couplings: $ Y_*^u\in(1,2)$ and $ Y_*^d \in(5,6)$ (bottom plot). We
find that more points satisfy both bounds than in the previous two cases.
This is expected since small $ Y_*^u$ suppresses one of the
contributions to
$\Gamma(b\rightarrow s \gamma)$\footnote{There
is also a contribution $\propto Y_d$ only as discussed
in section \ref{bsgonegen}.} while larger $Y_*^d$ suppresses
contribution to $\hbox{Im}\,C_{4K}$. However, the bound from
non-universal $Z\bar{b}_L b_L$ coupling correction is more
constrained in this case due to the
$(\frac{Y_*^{di3}}{Y_*^{u33}})^2$ enhancement in $\delta g_{Z\bar
b_L b_L}$ (see Eq. (\ref{Eqzbb})) so that we have to study the
consequence of this bound. In Fig. $\ref{zbbek5}$, we present the
result from the scan for $\hbox{Im}\,C_{4K}$ and $\delta
g_{Z\bar{b_L}b_L}$. We can see that when $Y_*^d = 5\sim 6$ and
$Y_*^u = 1\sim 2$ (right plot) the $\delta g_{Z\bar{b_L}b_L}$ bound
eliminates a majority of the points. However, for $Y_*^u = Y_*^d \in (3,4)$
(left plot), the bound on $\delta g_{Z\bar{b_L}b_L}$ is easily
satisfied, as expected from our analysis in Section \ref{Zbb}.
%
%

\begin{figure}
\begin{center}
\includegraphics[scale = 0.7]{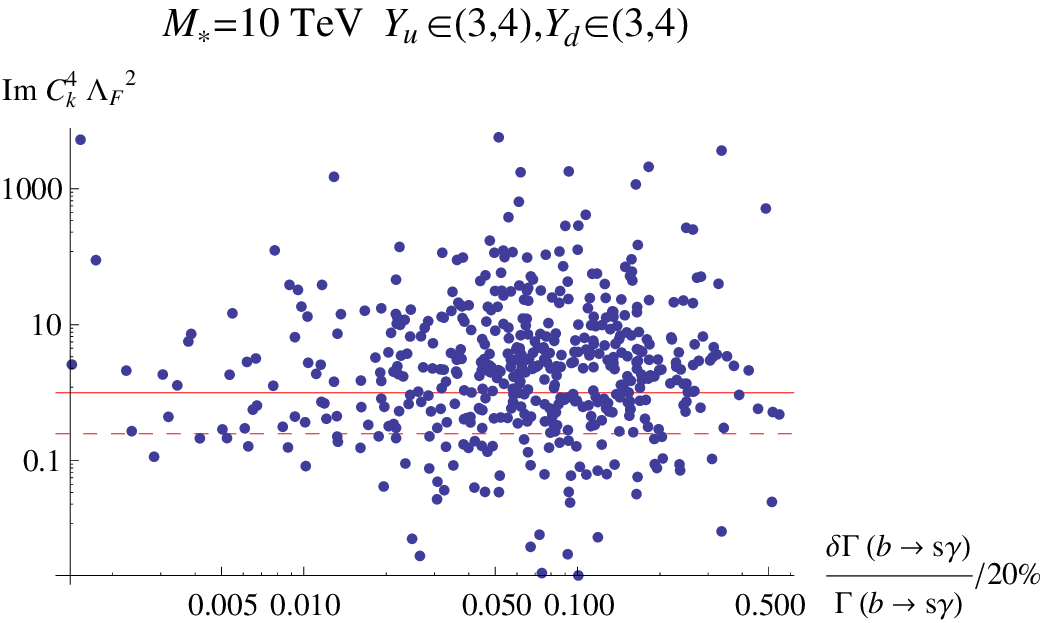}
\includegraphics[scale = 0.7]{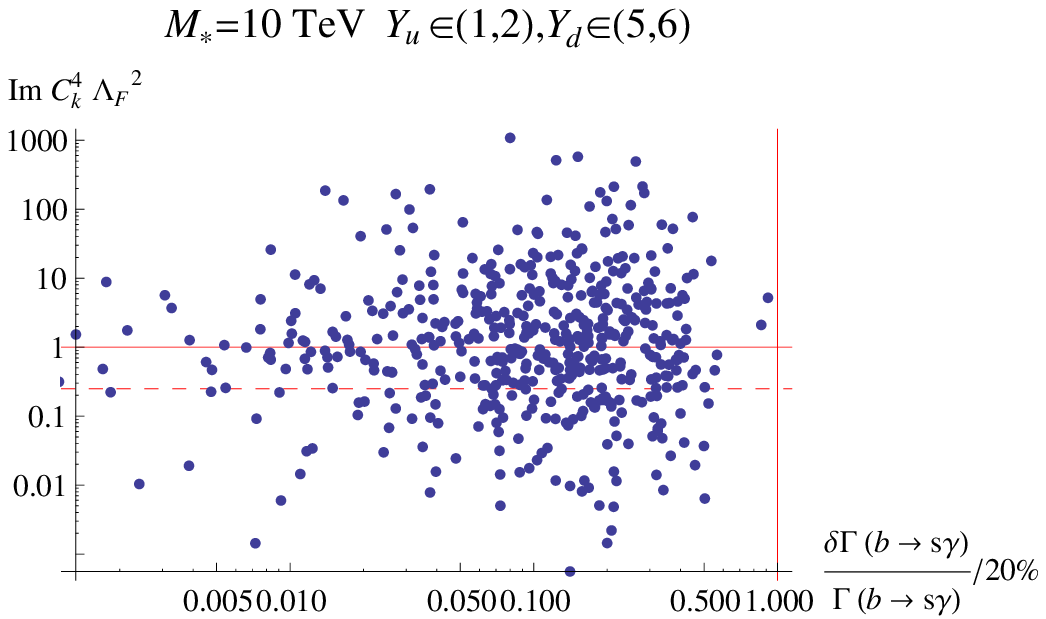}
\includegraphics[scale = 0.7]{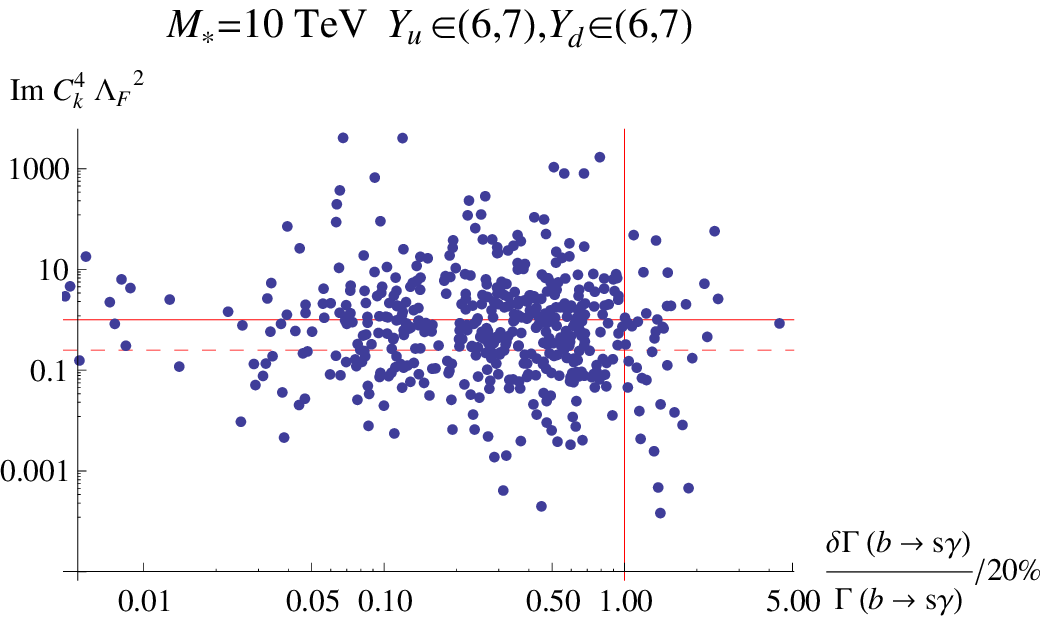}

\caption{\label{ekbsgm10} Same as Fig. \ref{ekbsgm5}, but with $M_* = 10\
\hbox{TeV}$.}
\end{center}
\end{figure}

\begin{figure}
\begin{center}
\includegraphics[scale = 0.7]{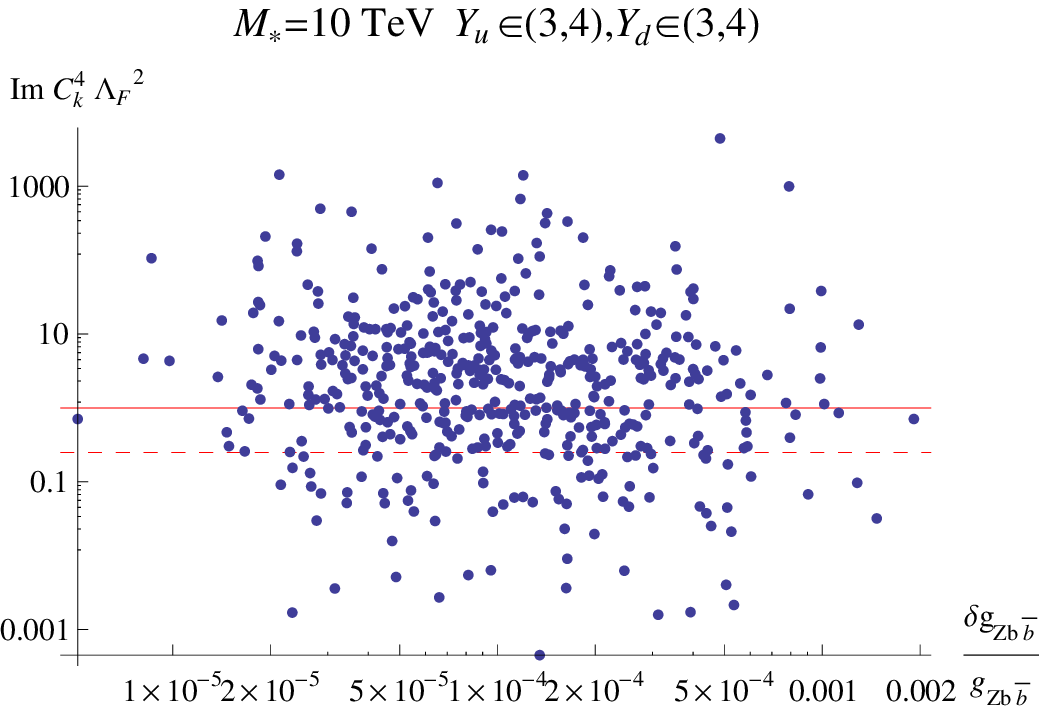}
\includegraphics[scale = 0.7]{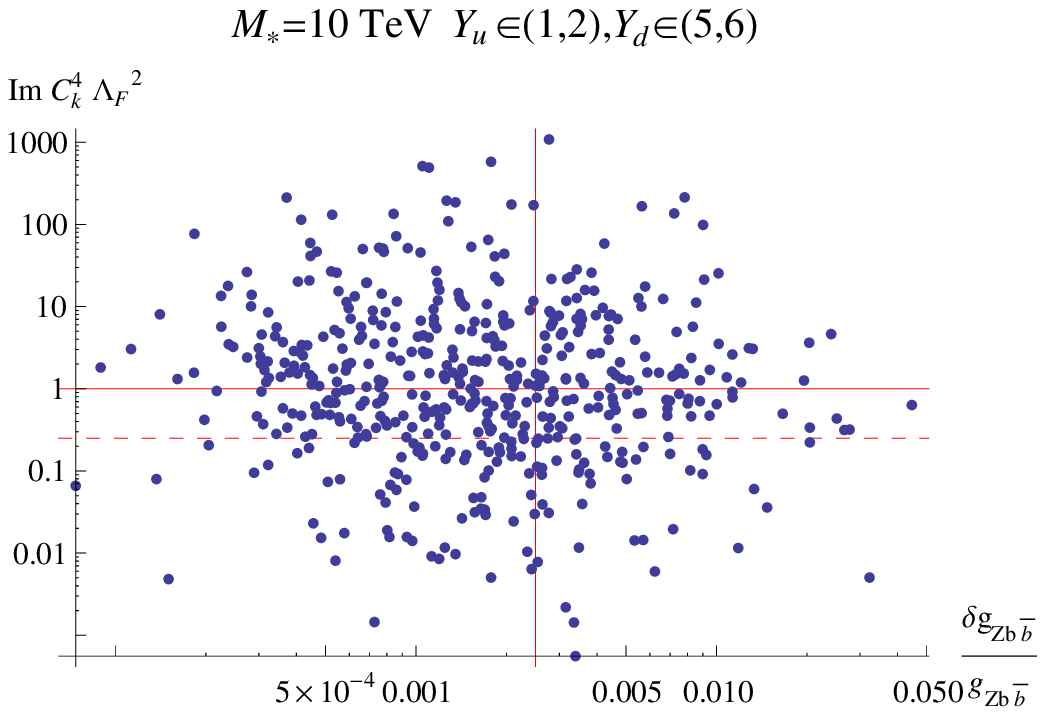}
\caption{\label{zbbek10}Same as Fig. \ref{zbbek5}, but with  $M_* = 10\
\hbox{TeV}$.}
\end{center}
\end{figure}

\begin{figure}
\begin{center}
\includegraphics[scale = 0.7]{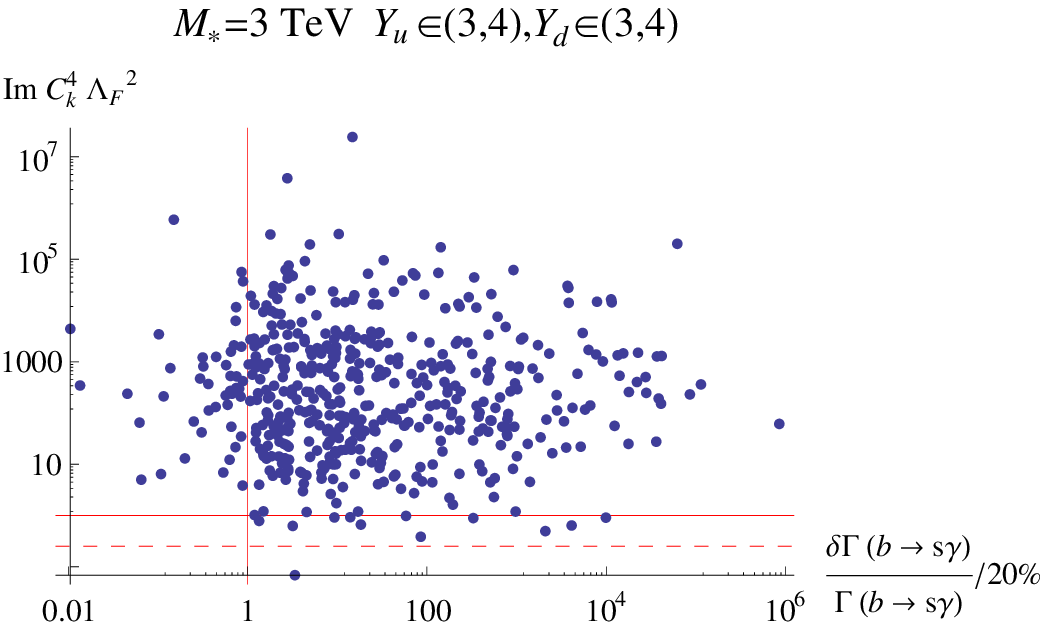}
\includegraphics[scale = 0.7]{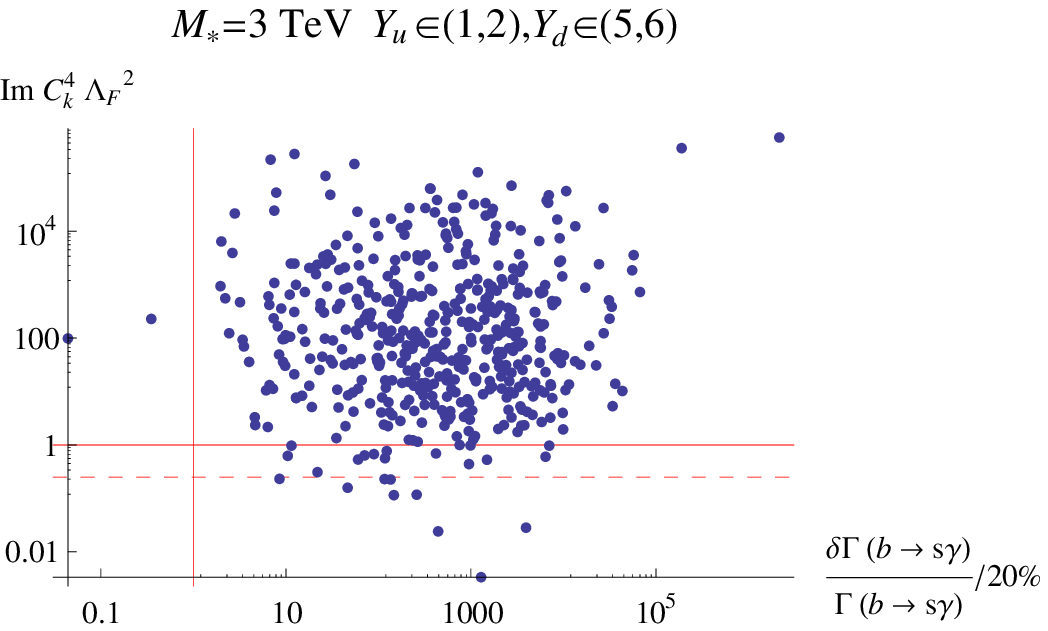}
\includegraphics[scale = 0.7]{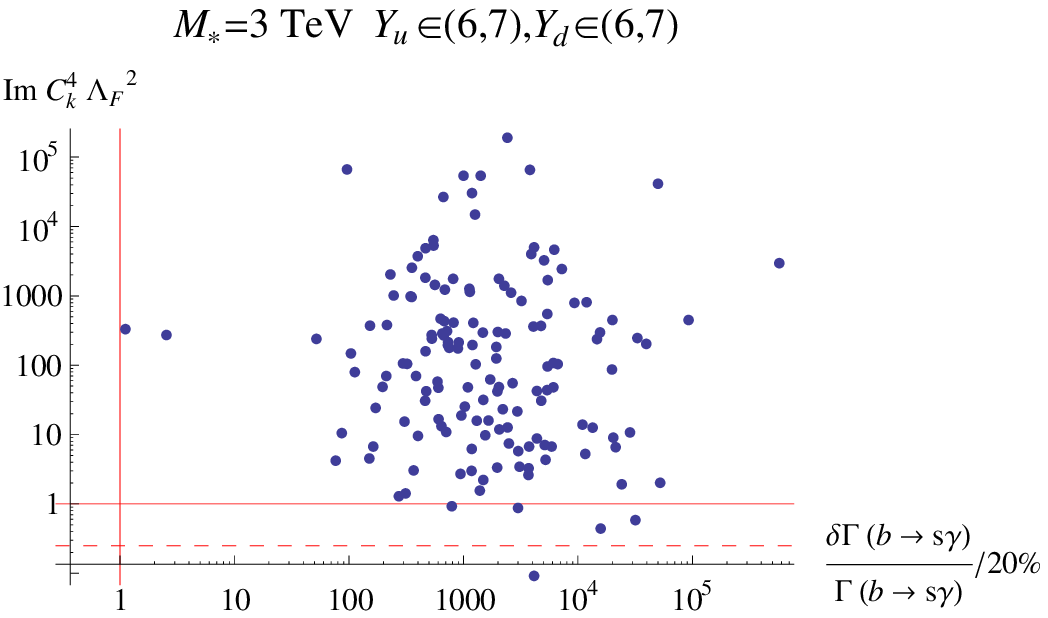}
\caption{\label{ekbsgm3} Same as Fig. \ref{ekbsgm5}, but with  $M_* = 3\
\hbox{TeV}$.}
\end{center}
\end{figure}

\begin{figure}
\begin{center}
\includegraphics[scale = 0.7]{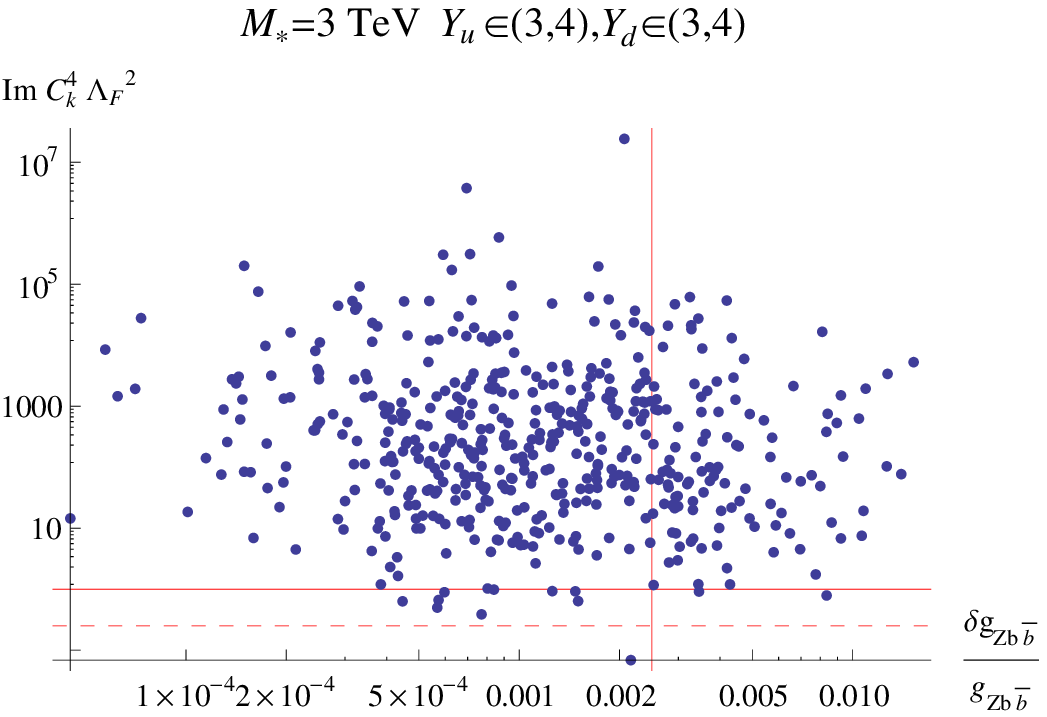}
\includegraphics[scale = 0.7]{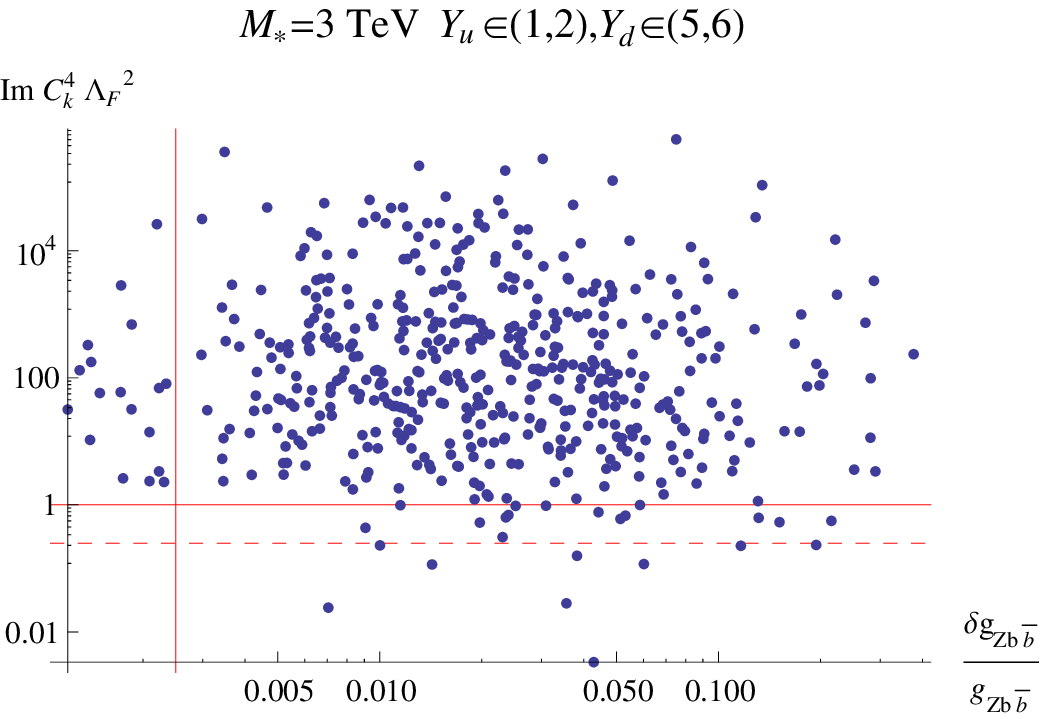}
\caption{\label{zbbek3}Same as Fig. \ref{zbbek5}, but with  $M_* = 3\
\hbox{TeV}$.}
\end{center}
\end{figure}

We show the same scatter plots for $M_* = 10\, \hbox{TeV}$ (Fig.
\ref{ekbsgm10}, \ref{zbbek10}) and $M_* = 3\,\hbox{TeV}$ (Fig.
\ref{ekbsgm3}, \ref{zbbek3}). As it is clearly shown in the plots,
all bounds can be easily satisfied for $M_* = 10\,\hbox{TeV}$, while
almost no point satisfy all bounds for $M_* = 3 \, \hbox{TeV}$. Note
that, with our choices of $Y_{ \ast }$, higher-order loop diagrams
with these couplings will give us corrections to all our observables
of $\sim Y_{ \ast }^2 / \left( 16 \pi^2 \right) \sim O
\left( 1 /
\hbox{a few} \right) - 1/ 10 $, which is the main source of error in our
analysis\footnote{Of course, we are also 
incurring an error of similar size due to neglect of higher KK modes
in the two-site approach for analyzing the $5D$ model}.

Now we consider the case with a larger
composite site gluon coupling, i.e.,
$g_{s*} = 6$. The
contribution in the
two-site model to $\Gamma(b \rightarrow s \gamma)$ is the same
as in the case $g_{s*} = 3$
while $\hbox{Im}\,C_{4K}$ increases by a factor of 4. Thus, rather than
showing separate plots for $g_{s*} = 6$, we can present the
{\em bounds}
for this case
on the same plots as for $g_{s*} = 3$ by just moving the line
from the
$\hbox{Im}\,C_{4K}$ bound downward
by factor of $4$. So all the points satisfying both constraints for
$g_{s*} = 6$ are below the {\em dashed} line and to the left of the solid
line in the same plots. As expected, for $g_{ s* } = 6$,
few (a sizable fraction of)
points satisfy the bounds for $M_* = 5 (10)$ TeV.

Combining the results of the
numerical analysis shown in the plots
with our earlier estimate in Eqs. (\ref{bounds})
and (\ref{Zbbbound}) of
$\sim 4.5$ TeV as the lowest heavy SM partner mass scale allowed,
we then conclude $M_*$ as low as
$\sim O(5)$ TeV with $g_{ \ast } \sim
3$ can satisfy all the constraints we considered.

\section{Conclusions}\label{conclusion}

The warped extra dimensional framework with bulk SM is very
well-motivated scenario for beyond the SM since it can address many
of the puzzles of nature. The two-site model provides a economical
%
%
description of this framework by effectively restricting to the SM
fields and their first KK excitations. In this paper, we studied
constraints on this model from flavor violation in the quark sector,
in particular, we showed that $\epsilon_K$ and BR $\left( b
\rightarrow s \gamma \right)$ provide the strongest constraints.
Moreover, these two observables have opposite dependences on the
composite site Yukawa couplings so that this parameter cannot be
used to ameliorate the flavor constraints.

Assuming anarchic composite site Yukawa couplings
and based on both numerical and analytical calculations, we showed that
$\sim O(5)$ TeV mass scale for the heavy states can be consistent
with both the observables for a size of composite site QCD
coupling which is consistent with the $5D$ AdS model
solving the
Planck-weak hierarchy problem, where the
$5D$ QCD coupling is matched to the
$4D$ coupling at the loop-level and with negligible
brane kinetic terms.
We argue that a larger
$5D$ coupling might be constrained by the requirement of $5D$ perturbativity.

Moreover, we showed that couplings in
the the two-site model are similar to those in
the $5D$ models with {\em bulk} Higgs (but
still leaning towards the TeV brane) rather than to the
brane-localized Higgs case. Thus our results suggest $\sim O(5)$ TeV
KK mass scale might be consistent with quark sector flavor violation
even for $5D$ AdS models with bulk Higgs.

With $O(5)$ TeV mass scale, signals at the LHC
(including its luminosity upgrade, the SLHC) from direct production of the
heavy states are extremely suppressed, in particular, only the heavy
gluon might be (barely) accessible \cite{kkgluon}. However, with very mild
tuning (i.e., deviation from anarchy in the composite site Yukawa
couplings), it is clear that $\sim O(3)$ TeV KK scale might be
allowed, enhancing the direct LHC signals and making even the EW
heavy states possibly accessible at the LHC \cite{Agashe:2007ki}.

Finally, we comment on future signals and constraints from flavor
violation. Obviously, the two-site model with $O(5)$ TeV mass scale
for the heavy particles is on the edge of $\epsilon_K$ and BR
$\left( b \rightarrow s \gamma \right)$ so that reduction of
theoretical errors in these observables will provide even stronger
constraints on this framework. More broadly speaking, given $\sim
O(5)$ TeV scale for the new particles, their contributions to
amplitudes for $\Delta F = 2$ and $\Delta F=1$ processes are
typically at the level of $\sim O(10 \%)$ of the SM.
Such a modest size might be relevant for the hints (i.e., $2-3
\sigma$ discrepancies in the SM unitarity triangle fit) of beyond-SM
effects in $B/K$ physics which are precisely at this level: see
references \cite{Lunghi:2007ak} for a general analysis and reference
\cite{Blanke:2008zb} for some discussion in the context of warped extra
dimension.

In fact, this size of new amplitudes in the warped framework can
still lead to striking deviations from SM, in spite of the (rough)
consistency of the SM predictions with the current flavor data.
For example, CP violation in $B_s$ mixing is expected to be $\sim
O(10 \%)$ in this scenario, larger than the SM expectation of a few
$\%$ (and there might be some hint for such an effect in the data
\cite{Bona:2008jn}).
This prediction can be thoroughly tested at the LHC-b.

An even more dramatic example is that a slightly larger
amplitude in $b \rightarrow s \gamma$, namely, $\sim O(0.5)$ of SM
with {\em opposite} chirality to that in the SM is still allowed since
such a contribution does not interfere with the SM amplitude {\em in
the decay width or BR}, giving an effect in BR $\sim O(20 \%)$ which
is on the edge of the current constraints. We discussed that in the
two-site model, the dominant effect in $b \rightarrow s \gamma$ is
precisely of such a size in the opposite-to-SM chirality amplitude.
The point is that such an effect gives $\sim O(0.5)$ time-dependent
mixing-induced CP asymmetry from the interference with the SM
amplitude, in sharp contrast to the SM prediction of this CP
asymmetry: $m_s / m_b \sim$ a few $\%$ (as already pointed out in
\cite{Agashe:2004ay, Agashe:2004cp} for the
$5D$ models). The current $2 \sigma$ error
on this CP asymmetry is $\sim O(0.5)$ so that there is no strong
constraint at present from this observable. However, if in the
future, this CP asymmetry can be probed at the SM level or even
$\sim O(0.1)$, then we might obtain a striking signal for this model
(or an even stronger constraint than $\sim O(5)$ TeV on the mass of
new particles). We leave this analysis for a future study.

\section*{Acknowledgments}

KA was supported in part by
NSF grant No. PHY-0652363
and would like to thank Csaba Csaki, Adam Falkowski,
Tony Gherghetta, Seung Lee,
Takemichi Okui, Michele Papucci, Gilad Perez, Raman Sundrum, Tomer
Volansky and Andreas Weiler
for discussions
and the Aspen Center for Physics for hospitality during part of this work.
The authors also thank Csaba Csaki, Adam Falkowski, Gilad Perez
and Andreas Weiler for comments on the manuscript.


\appendix

\section{Model Independent Loop Calculation}\label{A}

\begin{figure}
\begin{center}
\includegraphics[]{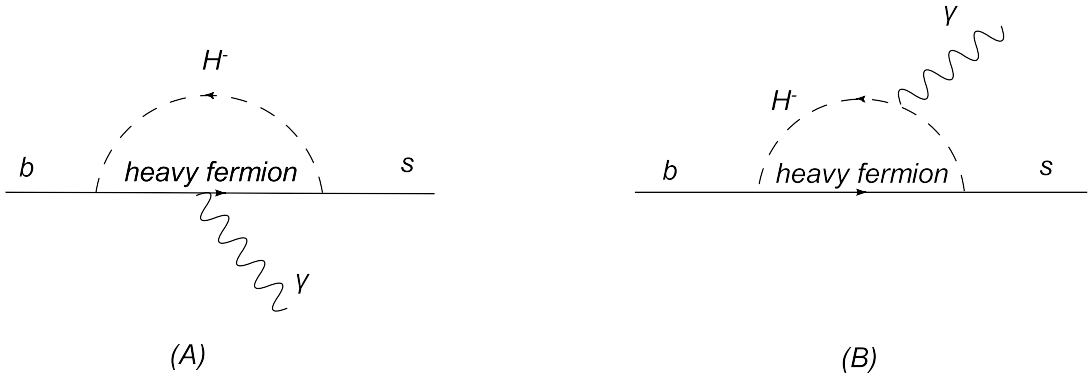}
\caption{\label{fig_feydiag}  Feynman diagrams for b $\rightarrow$ s
$\gamma$ via charged Higgs}
\end{center}
\end{figure}

We work in {\em non}-unitary gauge for the
electroweak gauge sector of the SM, where we must include the would-be
Goldstone bosons in the loop. The model-{\em independent} interaction
between a {\em charged} Higgs, SM down-type quarks ($d$) and
an up-type heavy quark ($U$) can be parametrized
as follows:
\begin{equation}
\label{alphas} \mathcal{L} \supset \bar{U}[\alpha_{1i} (1+\gamma_5)
+ \alpha_{2i}(1-\gamma_5)]d_i\  H^{-} + h.c.
\end{equation}
where {\em all quarks are in mass
eigenstate basis (including effects of EWSB)}. We focus on the
dominant contributions to the
dipole moment operator for $b \rightarrow s \gamma$
generated
by these interactions -- the relevant diagrams contain the charged
Higgs and heavy fermion in the loop with the
SM fermions as external legs (see Fig. \ref{fig_feydiag}A and
B). We will then apply the results obtained in this section for the
{\em specific} case of the two-site model and
calculate the effective dipole operator for {\em one} generation
in Appendix \ref{B} and $b \rightarrow s \gamma$ in appendix \ref{threegen}.

For the first diagram (see Fig. \ref{fig_feydiag}A), with photon
line attached to the heavy fermion, we get the effective
operator\footnote{We used Feynman gauge in this calculation.
Since we are considering only the dominant diagrams here,
the result will be different by
$O(\frac{M_w^2}{M_*^2})$
if we
use another non-unitary gauge.
Such differences can be neglected for our purpose here.
Of course including the other diagrams (with $W/Z$)
will produce a gauge-invariant result.
}
\begin{equation}
\label{loop1} \mathcal{H}^{\text eff}_1 = \frac{i e Q_U}{8\pi^2}
\frac{(2 \epsilon \cdot p)}{M_w^2}\left\{ A_1 \bar{s}(1+\gamma_5)b +
B_1 \bar{s} (1-\gamma_5)b \right\}
\end{equation}
with
\begin{eqnarray}
\label{loop2} A_1=(\alpha_{2b}\alpha_{2s}^* m_b +
\alpha_{1b}\alpha_{1s}^* m_s)f_1(t) +(\alpha_{2s}^*\alpha_{1b}) M_*
f_2(t) \\ \nonumber B_1 =(\alpha_{1b}\alpha_{1s}^*
m_b+\alpha_{2b}\alpha_{2s}^* m_s )f_1(t) +(\alpha_{1s}^*\alpha_{2b})
M_* f_2(t)
\end{eqnarray}
and
\begin{eqnarray}
\label{loop3} f_1(t) = - \frac{t[t(t-6)+3] + 6t\ln(t)
+2}{12(t-1)^4};\qquad f_2(t) = -\frac{(t-4)t +2\ln(t) + 3}{2(t-1)^3}
\end{eqnarray}
where $M_*$ is the mass of the heavy fermion, $Q_U$ is the charge of
the heavy fermion, $t = M_*^2/M_w^2$. This result can also be used
for the diagram with {\em neutral} Higgs (including
physical Higgs and the neutral would-be-Goldstone boson) in the loop.

The result for the second diagram(See Fig. \ref{fig_feydiag} B),
with photon attached to the charged Higgs, is
\begin{equation}
\label{loop4} \mathcal{H}^{\text eff}_2 = \frac{-ie }{8\pi^2}
\frac{(2 \epsilon \cdot p)}{M_w^2} \left\{ A_2\bar{s}(1+\gamma_5)b +
B_2\bar{s}(1-\gamma_5)b \right\}
\end{equation}
with
\begin{eqnarray}
\label{loop5} A_2 = (\alpha_{2b}\alpha_{2s}^* m_b+
\alpha_{1b}\alpha_{1s}^* m_s) g_1(t) + (\alpha_{1b}\alpha^*_{2s})
M_* g_2(t) \\ \nonumber B_2 =(\alpha_{1b}\alpha_{1s}^*
m_b+\alpha_{2b}\alpha_{2s}^* m_s) g_1(t) +
(\alpha^*_{1s}\alpha_{2b}) M_* g_2(t)
\end{eqnarray}
and
\begin{eqnarray}
\label{loop6} g_1(t) = \frac{2t^3 - 6t^2 \ln(t) - 6t + 1 +
3t^2}{12(t-1)^4};\qquad g_2(t) = \frac{t^2 - 2t \ln(t) -
1}{2(t-1)^3}
\end{eqnarray}
These results (Eq. \ref{loop1} and \ref{loop4}) can be applied to
calculate $\Gamma(b\rightarrow s\gamma)$ if we find the couplings
$\alpha_{1i}$, $\alpha_{2i}$ (see Eq. \ref{alphas}).

\section{Mass matrix diagonalization and dipole moment operator for one generation}\label{B}

Having performed a
calculation of the dipole
operator for $b \rightarrow s$ generated by general couplings
of bottom and strange quarks to Higgs and heavy fermions,
we now consider this contribution specifically in the two-site model.
As explained in
section \ref{bsgonegen}, we have
to
%
%
consider the mixing between the SM and heavy fermions of all
three generations
induced after EWSB. Diagonalization of this mixing will
give the couplings to Higgs in mass eigenstate basis for
the quarks which we can then plug into the model-independent results of
appendix \ref{A} in order to calculate $b \rightarrow s
\gamma$.
%
%
%
%
In this section, we will first consider {\em analytically}
the simpler {\em one} generation case, i.e.,
a calculation of $\left( g - 2 \right)_{ \mu }$,
which will be generalized (numerically) to the case of three
generations for calculating $b \rightarrow s \gamma$
in the next sub-section.
This result for the dipole operator for one
generation was also used in section \ref{bsgonegen} to obtain an
{\em estimate}
for $b \rightarrow s \gamma$ (after multiplying by an estimate
for the generational mixing factors).
%

%
%

The one generation mass matrix for down type quarks (including
effects of EWSB) is (see Eq.
(\ref{yukawa}))
\begin{equation}
\label{mass} (\bar{b}_L\quad \bar{\tilde{B}}_L\quad \bar{B}_L)M_*
\left(
\begin{array}{ccc} x s_{q} s_{b} & 0 &  x s_{q} \\
0 & x & 1 \\
x s_{b} & 1 & x  \end{array} \right)\left(\begin{array}{c}
\tilde{b}_R \\ B_R \\ \tilde{B}_R \end{array}\right) + h.c.
\end{equation}
where $x = vY_* / \left( M_*\sqrt{2} \right)$, $\tilde{B}$ and $B$ are
composite $SU(2)_L$ singlet and doublet fermions respectively. It
can be diagonalized by bi-unitary transformation to first order in
$x$.
\begin{equation}
O_{D_L} = \left( \begin{array}{ccc}  1 & x s_{q}/\sqrt{2} & - x
s_{q}/\sqrt{2} \\ - x s_{q} & 1/\sqrt{2} & -1/\sqrt{2} \\ 0 &
1/\sqrt{2} & 1/\sqrt{2}
\end{array} \right); \qquad O_{D_R} = \left( \begin{array}{ccc}  1 & x s_{b}/\sqrt{2} &  x
s_{b}/\sqrt{2} \\ - x s_{b} & 1/\sqrt{2} & 1/\sqrt{2} \\ 0 &
1/\sqrt{2} & -1/\sqrt{2}
\end{array} \right)
\end{equation}

\begin{equation}
O_{D_L}^\dagger \left(
\begin{array}{ccc} x s_{q} s_{b} & 0 &  x s_{q} \\
0 & x & 1 \\
x s_{b} & 1 & x  \end{array} \right) O_{D_R} = \hbox{diag}
(\ x s_q s_b \ ,
1+x\ , 1-x\ )
\end{equation}
Similarly we can get the up-type diagonalization matrix $(O_{U_L})$
and $(O_{U_R})$. We define the mass eigenstates as
\begin{equation}
\left(\begin{array}{c} b_L^{\text SM} \\ B_{1L} \\ B_{2L}
\end{array} \right) = O_{D_L}^\dagger \left(
\begin{array}{c} b_L \\ \tilde{B}_{L} \\ B_{L}
\end{array} \right) \qquad
\left(\begin{array}{c} b_R^{\text SM} \\ B_{1R} \\ B_{2R}
\end{array} \right) = O_{D_R}^\dagger \left(
\begin{array}{c} \tilde{b}_R \\ B_{R} \\ \tilde{B}_{R}
\end{array} \right),
\end{equation}
where $b^{\text SM}$ is the SM bottom quark with mass $v Y_* s_{q}
s_{b}$. $B_1$ is the heavy state with mass $(1+x)M_*$ and $B_2$ is
the heavy state with mass $(1-x)M_*$. Similar mass eigenstates can
be defined for up-type quarks $(t^{\text SM}, T_1, T_2)$.

The coupling between down type and up type quarks through charged
Higgs is
\begin{equation}
Y_*(\bar{b^{\text SM}_L} \quad \bar{B}_{1L} \quad \bar{B}_{2L})\
O_{D_L}^\dagger \left( \begin{array}{ccc} s_{q} s_{t} & 0 & s_{q}
\\ 0 & -1 & 0
\\ s_{t} & 0 & 1
\end{array} \right) O_{U_R} \left( \begin{array}{c}  t^{\text SM}_R \\
T_{1R} \\ T_{2R}  \end{array} \right) H^-
\end{equation}
We can find the couplings between $b^{\text SM}_L$ and heavy up-type
quarks
\begin{equation}
Y_*H^- \bar{b}^{\text SM}_L \left[ \frac{(1+x)}{\sqrt{2}} s_{q}
T_{1R} + \frac{(x-1)}{\sqrt{2}} s_{q} T_{2R} \right]
\end{equation}
Similarly, we have the coupling coming from another chirality
\begin{equation}
Y_* (\bar{b}^{\text SM}_R \quad \bar{B}_{1R} \quad \bar{B}_{2R}   )\
O_{D_R}^\dagger \left(\begin{array}{ccc} -s_{q} s_{b} & 0 & - s_{b} \\
0 & 1 & 0 \\
- s_{q} & 0 & -1\end{array} \right) O_{U_L} \left( \begin{array}{c} t^{\text SM}_L \\
T_{1L} \\ T_{2L}
\end{array} \right)H^-
\end{equation}
which gives us the coupling
\begin{equation}
Y_*H^- \bar{b}^{\text SM}_R \left[ -\frac{(1+x)}{\sqrt{2}} s_{b}
T_{1L} + \frac{(x-1)}{\sqrt{2}} s_{b} T_{2L}  \right]
\end{equation}
Altogether, we have the charged Higgs coupling between SM bottom
quark and heavy up-type quark
\begin{eqnarray}
Y_*H^-\bar{b}^{\text SM} \left[ (\frac{1+\gamma_5}{2})(1+x)
\frac{s_{q}}{\sqrt{2}} -
(\frac{1-\gamma_5}{2})(1+x)\frac{s_{b}}{\sqrt{2}} \right]T_1 + \\
\nonumber Y_*H^-\bar{b}^{\text SM} \left[ (\frac{1+\gamma_5}{2})
\frac{x-1}{\sqrt{2}} s_{q} + (\frac{1-\gamma_5}{2})
\frac{x-1}{\sqrt{2}}s_{b} \right] T_2
\end{eqnarray}
Based on our parametrization of the couplings (see Eq.
\ref{alphas}), we extract (we ignore the subscript ``b'' in
$\alpha_{1,2}$ here)
\begin{eqnarray}
\label{alpha1gen}
\alpha_1^{(1)}&=& - \frac{(1+x) s_{b}}{2\sqrt{2}}Y_*\\ \nonumber
\alpha_1^{(2)}&=& \frac{(x-1) s_{b}}{2\sqrt{2}}Y_*\\ \nonumber
\alpha_2^{(1)} &=&\frac{(1+x) s_{q}}{2\sqrt{2}}Y_*\\ \nonumber
\alpha_2^{(2)} &=& \frac{(x-1) s_{q}}{2\sqrt{2}}Y_*
\end{eqnarray}
The contribution from heavy up-type quark to the dipole moment
operator would be (see Eq. \ref{loop1} and \ref{loop4})
\begin{equation}
\mathcal{H}^{\text{dipole}}_{\text{charged Higgs}} =  \frac{i e
}{8\pi^2} \frac{(2 \epsilon \cdot p)}{M_w^2} K \left[ \bar{b}^{\text
SM} (1-\gamma_5) b^{\text SM}+ \bar{b}^{\text SM} (1+\gamma_5)
b^{\text SM} \right]
\end{equation}
with
\begin{eqnarray}
\label{K}
K = \sum_{i=1}^2 \left(|\alpha_1^{(i)}|^2 +
|\alpha_2^{(i)}|^2\right) m_b\left[\frac{2}{3} f_1(t_i)-
g_1(t_i)\right] +\sum_{i=1}^2 (\alpha_1^{(i)*}\alpha_2^{(i)} )
M_i\left[\frac{2}{3} f_2(t_i)-g_2(t_i)\right]
\end{eqnarray}
Substituting Eq. (\ref{alpha1gen}) in (\ref{K}) one can see that the first term is sub-leading due to additional powers of $s_b$,$
s_q$. For the second term we use the approximation
\begin{equation}
\frac{2}{3} f_2(t_i)-g_2(t_i) \approx
-\frac{5}{6}\frac{M_w^2}{(M_*)^2(1\pm x)^2}
\end{equation}
It gives us
\begin{equation}
K \approx \frac{5x}{24}s_{q} s_{b} \frac{M_w^2}{(M_*)^2} (Y_*)^2
\end{equation}
And the final result is
\begin{equation}\label{Eq.charged Higgs}
\mathcal{H}^{\text{dipole}}_{\text{charged\,Higgs}} =
\frac{5}{12}(Y_*)^2m_b\frac{i e}{16 \pi^2} \frac{(2\epsilon\cdot
p)}{(M_*)^2} [ \bar{b}^{\text SM} (1-\gamma_5) b^{\text
SM}+\bar{b}^{\text SM} (1+\gamma_5) b^{\text SM}]
\end{equation}
Note that we have chosen {\em not} to combine the two terms in $\Big[...\Big]$
in the above equation. The reason is that
when we apply the above result to $b \rightarrow s \gamma$,
then the two terms with {\em different} chirality structure will be multiplied
by different mixing angles and hence it is useful to keep
track of the two terms separately even for the case of one generation.

The contribution from neutral Higgs can be calculated in a similar
fashion. The coupling between down-type quarks and neutral Higgs is
\begin{equation}
Y_* H^0 \left( \bar{b}^{\text SM}_L \quad \bar{B}_{1L} \quad
\bar{B}_{2L}
\right) O_{D_L}^\dagger \left( \begin{array}{ccc} s_q s_b & 0 & s_q \\
0 & 1 & 0 \\ s_b & 0 & 1
\end{array}\right) O_{D_R} \left( \begin{array}{c} b^{\text SM}_R \\ B_{1R} \\ B_{2R}
\end{array}\right) + h.c.
\end{equation}
From this we can find the coupling between SM $b$ quark and heavy
down-type fermions:
\begin{eqnarray}
Y_* H^0 \left\{ \bar{b}^{\text SM}_L \left[ \frac{1-x}{\sqrt{2}} s_q
B_{1R} - \frac{1+x}{\sqrt{2}} s_q B_{2R} \right]   +  \bar{b}^{\text
SM}_R \left[ \frac{1-x}{\sqrt{2}} s_b B_{1L} +
\frac{1+x}{\sqrt{2}}s_b B_{2L} \right] \right\} + h.c.
\end{eqnarray}
which gives us (see Eq. \ref{alphas})
\begin{eqnarray}
\alpha^{(i)}_1 = Y_* \frac{s_b}{2\sqrt{2}} (1-x, \; 1+x)\\
\nonumber \alpha_2^{(i)} = Y_* \frac{s_q}{2\sqrt{2}} (1-x,\; -1-x)
\end{eqnarray}
Follow the same procedure as before, including only the first
diagram (Fig. \ref{fig_feydiag}A). We get
\begin{equation}
\label{Eq.neutral Higgs}
\mathcal{H}^{\text{dipole}}_{\text{neutral\,Higgs}} =
-\frac{1}{4}(Y_*)^2m_b\frac{i e}{16 \pi^2} \frac{(2\epsilon\cdot
p)}{(M_*)^2} [ \bar{b}^{\text SM} (1-\gamma_5) b^{\text
SM}+\bar{b}^{\text SM} (1+\gamma_5) b^{\text SM}]
\end{equation}

\subsection{Three generation calculation}
\label{threegen}

Generalizing to three generations, the mass matrix Eq. (\ref{mass})
becomes $9\times 9$. However, since analytical diagonalization of
this $9\times 9$ matrix is difficult, we do it numerically and
extract the parameters $\alpha_1,\alpha_2$ (see Eq. (\ref{alphas})
which parametrize general interaction between fermions and Higgs
field, keeping in mind that $\alpha_{1,2}$ will now have six
components $\alpha_{1,2}^{(1,2,...6)}$ because we have six heavy
mass eigenstates). Then using these $\alpha$'s in the formulae from
the loop calculation in Eqs. (\ref{loop1}) and (\ref{loop4}), we
will get {\em exact} values for the $C_7$ and $C_7^\prime$
coefficients in the amplitude for $b \rightarrow s \gamma$ (instead
of the estimates presented in section \ref{bsgonegen}). Similarly,
applying the above diagonalization to Eq. (\ref{gauge}) allows us to
calculate the flavor-violating couplings of heavy gluon to the SM
fermions {\em after} EWSB (including effects of SM-heavy fermion
mixing) which generate contributions to $\epsilon_K$.
The results of the numerical scan in section \ref{numerical} are
based on these calculations.

\section{Details of Scan}\label{scan}

All the masses and mixings in the fermion sector
(including SM and heavy) can be
parametrized by the composite site
Yukawa couplings ($Y^*_{u,d}$) and the elementary/composite
mixings ($s_{q,d,u}$). Of course, we must choose
$Y_*$ and $s_{q,u,d}$ to give the observed quark masses and CKM angles.
We would like the composite site Yukawa couplings to be
``anarchical'', i.e., of the same order, and $s_{q,d,u}$ to be
hierarchical\footnote{As mentioned earlier, these
assumptions can be justified by the
correspondence with the $5D$ model to be discussed later.} 
in order to explain SM fermion masses and mixing.
This anarchy
condition and Eqs. \eqref{mixing angles} and \eqref{quarkmass}
(with generalization to other quark masses) lead to
the following rough size of the mixing angles
\begin{eqnarray}
\label{sin} s_{u3} \sim 1, ~~~s_{d3} \sim \frac{s_{u3}m_b
Y^{u}_{*}}{m_t Y_{*}^{d}},~~~
s_{q3} \sim \frac{m_t \sqrt{2}}{v s_{u3}Y_{*}^{u}},\nonumber\\
s_{d2} \sim \frac{m_s s_{d3}}{ m_b\lambda^2 },~~~ s_{u2} \sim
\frac{s_{u3} m_c}{\lambda^2 m_t},~
~~s_{q2} \sim \lambda^2s_{q3},\nonumber\\
s_{d1}\sim\frac{m_d s_{d3}}{m_b\lambda^3},~~~ s_{u1}\sim\frac{m_u
s_{u3}}{m_t\lambda^3},~~~ s_{q1}\sim\lambda^3s_{q3}.
\end{eqnarray}
We choose to scan over the following independent variables
\begin {itemize}
\item
 Elementary-composite mixing angles $s_{q,u,d}$
\item
SM rotation matrices $U_R,U_L,D_R$ ($D_L$ is fixed by $D_L=U_L\cdot
V_{CKM}$)
\end{itemize}
(This choice is equivalent to treating $Y^{u,d}_*$ and $s_{q,u,d}$
as the independent variables which are scanned.)
We randomly vary each set of the independent variables around their
``natural'' size by a factor of three, where the natural sizes for
the $s_{q,u,d}$ are defined to be Eq. \eqref{sin} and that for
$U_R,U_L,D_R$ in Eq. \eqref{mixing angles} by replacing ``$\sim$''
by ``$=$'' in both these equations. 
Then we calculate corresponding $Y_{* u,d}$\footnote{We
are ignoring the mixing between the SM and heavy fermions induced by
EWSB in the last relation here
which results in an error of $Y_*^2 v^2 / M_*^2 \sim
\hbox{a few} \; \%$ (for the our choice of parameters) in the determination of
$Y_{ \ast }$.}
\begin{eqnarray}
&Y_u=\frac{\sqrt{2}}{v}(U_L)\cdot M_u^{diag}\cdot ({U_R})^{\dagger};
\qquad
Y_d=\frac{\sqrt{2}}{v}(D_L)\cdot M_d^{diag}\cdot {(D_R)}^{\dagger}\nonumber\\
&Y^{u,d}_{*} \approx s_Q^{-1} Y_{u,d}s_{u,d}^{-1}
\end{eqnarray}
Then we check whether our $Y_{*u,d}$ are ``anarchical'', i.e.,
whether they satisfy the following condition
\begin{eqnarray}
   \frac{\hbox{Max}(|Y_{*u}|)}{3} <\text{G.M.}(|Y_{*u}|)<3* \hbox{Min}(|Y_{*u}|)\nonumber\\
   \frac{\hbox{Max}(|Y_{*d}|)}{3} <\text{G.M.}(|Y_{*d}|)<3* \hbox{Min}(|Y_{*d}|)
\end{eqnarray}
where G.M. stands for the geometrical mean. If these Yukawas satisfy
``anarchy'' condition, we proceed to calculate new physics
contribution to $\Gamma(b \rightarrow s \gamma)$, Im$\,C_4^K$
(as described in section \ref{threegen})
and
$\delta g_{Z\bar{b}_L b_L}$ as in Eq. (\ref{Eqzbb}).
On the other hand, if these Yukawas do not satisfy the anarchy condition,
then we discard them.
%
%
We have checked
that the couplings ($Y_{*u,d}$) generated in this way are random, i.e.,
that there is no correlation between different elements of the
matrices. The results of the scan are presented in Fig.
\ref{ekbsgm5} to \ref{zbbek3}.

\section{Sub-leading effects}

\subsection{$\epsilon_K$}

Similarly to the heavy gluon exchange, heavy EW gauge boson
exchange generates $(V-A) \times (V+A)$ operators, but it gives
$C_5 \left( M_{ \ast } \right)$ only and of smaller size than $C_4
\left( M_{ \ast } \right)$ from heavy gluon due to smaller
values of gauge
couplings and gauge quantum numbers in the heavy EW boson exchange than in
heavy gluon exchange.
Moreover, the model-{\em independent} constraint from UTfit
\cite{Bona:2007vi} on $C_5 \left( \hbox{renormalized at a few TeV
scale} \right)$ is weaker than for $C_4$.
So, we find that constraint on $M_{ \ast }$ from heavy EW gauge
boson exchange in the two-site model is weaker than that from heavy
gluon exchange: see also discussion in \cite{Blanke:2008zb}.

We have also checked that the constraint from $(V-A) \times (V-A)$
and $(V+A) \times (V+A)$-type operators from heavy gauge boson
exchange in the two-site model can be weaker than from $(V-A) \times
(V+A)$ operator from heavy gluon exchange.
In detail, such exchange generates the Wilson coefficient $C_1
\left( M_{ \ast } \right)$. Firstly, the model-independent bound
on $C_1 \left( \hbox{renormalized at a few TeV scale} \right)$ is
weaker than for $C_4$
%
%
due to the absence of matrix element {\em and} RGE enhancement for
$C_1$ relative to $C_4$.
Secondly, in the two-site model, the size of $C_1$ can be
%
%
effectively {\em controlled} by a single parameter, namely, the
amount of elementary-composite mixing of $b_L$ -- the point being
that
the other down-type elementary-composite mixings are then fixed: the
ones for $d_L, s_L$ via CKM mixing angles and then, for given
composite Yukawa, the right-handed ones by SM Yukawa (as discussed
earlier).\footnote{Contrast this case to that for $C_{ 4, \; 5 }$
above whose size was {\em fixed} in terms of SM fermion Yukawa
couplings/masses (due to a combination of left and right-handed
elementary-composite mixings involved in $C_4$).}
Usually, one chooses $s_{ q3 }
%
%
$ to satisfy the
constraint from $Z b \bar{b}$ (as discussed earlier) and
simultaneously to obtain the correct top Yukawa, i.e., $Y_{ \ast }
s_{ q3 }
%
%
\sim O(1)$, assuming SM $t_R$ is fully
composite.
For the choice of $M_{ \ast } \sim $ a few TeV and $Y_{ \ast } \sim$
a few, we then find $
s_{ q3 }
%
%
\sim 1 /$ (a few).
With this size of $s_{ q3 }
%
%
$ and once we choose
$M_{ \ast }$ to satisfy the $\epsilon_K$-constraint from $(V-A)
\times (V+A)$ operators, we find that both $(V-A) \times (V-A)$ and
$(V+A) \times (V+A)$-type operators do not give as strong a
constraint as from heavy gluon contribution to the $(V-A) \times
(V+A)$ operator: see also \cite{Blanke:2008zb} for a related
discussion.

\subsection{Other $B$-physics observables}

It is easy to compute $B_{ d, \; s }$ mixing amplitudes in the
two-site model. The main new physics contribution comes from the
flavor violating couplings of heavy gluon, just like for $\Delta S =
2$ process discussed earlier. We have checked that bounds on $B_{ d,
\; s}$ mixing amplitude is satisfied once $\epsilon_K$ is safe: see
also \cite{Fitzpatrick:2007sa, Davidson:2007si,
Csaki:2008zd, Blanke:2008zb} for related
discussions.

In detail, the $(V-A) \times (V+A)$ type operator generated in the
two-site model is less constrained in the $B_{ d, \; s }$ systems
than in the $K$ system for the following reasons. Firstly, the
model-independent constraint on $C^{new}_{ 4, \; 5 } \left( M_\ast
\right) / C^{ SM }_1 \left( M_W \right)$ is weaker in the $B_{ d, \;
s }$ system than in the
$K$ system since there is no matrix element enhancement for $C_{ 4,
5 }$ in the $B_{ d, \; s }$ mixing operators (unlike for $K$
mixing). Secondly, in the two-site model, the size of $C^{new}_{ 4,
\; 5 } \left( M_{ \ast } \right) / C^{ SM }_1 \left( M_W \right)$
for $B_{ d, \; s }$ mixing turns out (due to the particular values
of down-type quark masses) to be smaller than in $K$ mixing. For the
$(V \pm A ) \times (V \pm A )$ type operator, the analysis is
similar to that for $K$ mixing.

Besides $\Delta F = 2$ processes, there are also new physics
contribution to $\Delta F = 1$ processes in the two-site model. For
example, the non-universal shift (in gauge eigenstate basis) in the
$Z$ couplings for $b_L$ (vs. $d_L$, $s_L$) will lead to
flavor-violating couplings to $Z$ once we transform to mass
eigenstate basis, resulting in the (flavor-violating) processes $b
\rightarrow s f \bar{f}$, where $f =$ quark, lepton. We have checked
that the new physics contribution to $b \rightarrow s l^+ l^-$
process is below the experimental bound once we satisfy $\delta
g_{Z\bar{b_L}b_L}/g_{Z\bar{b_L}b_L} \lesssim 0.25\%$ as required by
the flavor-preserving $Z b \bar{b}$ data: see also
\cite{Casagrande:2008hr} for a related discussion.

We also checked the new physics contribution to the time-dependent
CP asymmetry in $b \rightarrow s \gamma$, i.e., $S_{CP}$ which
requires an interference between the $C_7$ and $C^\prime_7$
amplitudes: $S_{CP}
%
%
\sim
C^\prime_7  C_7/
(|C_7|^2+|C_7^\prime|^2)$.
In the SM, $S_{CP} \sim m_s/m_b$ due to the suppression of
$C^\prime_7$ by $m_s/m_b$ relative to $C_7$ \cite{Atwood:1997zr}.
In the two-site model, new physics contribution will generically
give $C^\prime_7 \sim C^{ SM } _7$ so that
%
%
we expect $S_{CP}$ to be sizable in the two-site model. However, we
found that there is no significant constraint coming from $S_{ CP }$
because of the large experimental uncertainty at present
\cite{HFAG1}.

\section{Relation to $5D$ AdS Model}
\label{relationto5d}

The two-site model can be considered to be a deconstruction of the
$5D$ AdS model with the following metric:
\begin{eqnarray}
\left( d s \right)^2 & = & e^{ - 2 k y }
d x^{ \mu } d x _{ \mu } + \left( d y \right)^2\qquad  (0 \leq y \leq \pi R\nonumber) \\
 & \equiv & \frac{1}{ \left( k z \right)^2 } \;
\left( d x_{ \mu } d x ^{ \mu } + dz^2 \right) \;\qquad \left(z_h
\leq z \leq\ z_v \right)
\end{eqnarray}
where $z \equiv e^{ k y } / k$, and with the two endpoints $z_h
\equiv 1 / k$ and $z_v \equiv e^{ k \pi R } / k$ being denoted by
the Planck and the TeV branes, respectively. The curvature scale
$k$ is taken to be of the order of the $4D$ Planck scale
($M_{ Pl }$)  and we
choose $k \pi R \sim \log \left( M_{ Pl } / \hbox{TeV} \right)$
with the SM Higgs being localized near the TeV brane in order to
solve the Planck-weak hierarchy problem. The KK masses are
quantized in units of $\sim k e^{ - k \pi R } \sim$ TeV and are
localized near the TeV brane. The composite site of the two-site model
corresponds (roughly) to
the TeV brane and elementary site to the Planck brane (but moved
close to the TeV brane with renormalization of the interactions
localized on it, i.e., as per holographic RG flow \cite{holorg}).
The mass eigenstates of the two-site model {\em
before} EWSB (which are admixtures
of elementary and composite site particles) correspond to the zero
and KK modes in the $5D$ model\footnote{These modes
are non-vanishing on {\em both} the Planck and the
TeV branes, corresponding to the admixtures in the
two-site model.}
(again with Higgs vev
set to zero), i.e.,
\begin{eqnarray}
\hbox{SM states} &\leftrightarrow & \hbox{zero modes}\\ \nonumber
\hbox{heavy states}&  \leftrightarrow & \hbox{KK modes}
\end{eqnarray}
Here, we show the correspondence between the couplings of the
two-site and the $5D$ AdS model. For
more discussion on this issue, see reference \cite{Contino:2006nn}.

\subsection{Size of composite gauge coupling}
Since $g_*$ is the composite site coupling, i.e., the coupling of
composite sector ($\approx$ heavy) gauge bosons to the composite
sector ($\approx$ heavy) fermions, it should correspond (roughly) to
the coupling of single gauge KK mode to two KK fermions. We find
that this coupling in the $5D$ model is (roughly) given by $g_5
\sqrt{k}$, where $g_5$ is the dimensionful $5D$ gauge coupling. This
result is (almost) independent of the range of profiles of KK
fermions that is of interest for down-type quarks (see section
\ref{KKcoupling}).
%
%
Hence, we identify:
\begin{eqnarray}
g_* & \leftrightarrow & g_5 \sqrt{k}
\label{relationgauge}
\end{eqnarray}

\subsubsection{Matching to QCD coupling}
\label{matching qcd} The value of the 5D coupling $g_5 \sqrt{k}$ can
be fixed by matching it to the $4D$ QCD coupling \cite{Pomarol:2000hp,
Choi:2002ps}:
\begin{eqnarray}
\frac{1}{ g^{ QCD \; 2 } } & \approx & \log \left( M_{ Pl } /
\hbox{TeV} \right) \left( \frac{1}{ \left( g^{ QCD \; 2 }_5 k
\right) } + \frac{ b^{ QCD } }{ 8 \pi^2} \right) + \frac{1}{ g^{ QCD
\; 2 }_{ Planck } } + \frac{1}{ g^{ QCD \; 2 }_{ TeV } }
\label{match}
\end{eqnarray}
where $1 / g^{ QCD \; 2 }_{ Planck, \; TeV}$ are the bare/tree-level
(positive) brane-kinetic terms and the $b^{ QCD } \approx -7$ is the
result of one-loop running effects. So  using $g^{ QCD } \sim 1$
(renormalized at the TeV scale) we get
\begin{itemize}

\item
$g_5 \sqrt{k} \sim 3$ for matching at the {\em loop} level, i.e.,
including the $b^{ QCD }$ term with zero bare/tree-level brane
kinetic terms and with a Planck-weak hierarchy. Clearly, this is the
{\em smallest} allowed value of $g_{ \ast }$ for this hierarchy.

\item
$g_5 \sqrt{k} \sim 6$ for matching at the {\em tree}-level, i.e.,
neglecting the $b^{ QCD }$ term, with no brane kinetic
terms\footnote{Equivalently, choosing the tree-level brane
kinetic term to cancel the loop contribution'': see discussion in
\cite{Csaki:2008zd}.}.

\end{itemize}

In general, the value of 
$g_5 \sqrt{k}$ can be even larger than above if we allow
non-zero (positive) brane kinetic terms (on the Planck or TeV
brane). In particular, with non-zero {\em Planck} brane localized kinetic
terms, the couplings of (lightest) gauge KK are still set by $g_5
\sqrt{k}$ since these modes are localized near TeV brane. Thus, the KK
coupling (measured in units of SM gauge coupling) also increases as
these brane kinetic terms are increased.
On the other hand, allowing (sizable) {\em TeV} brane localized kinetic
terms has a more interesting effect as follows. The value of $g_5
\sqrt{k}$ (again measured in units of the SM gauge coupling)
increases as in the case of the Planck brane localized kinetic
terms, but the KK gauge coupling is clearly determined by the
kinetic term localized on the TeV brane where the KK modes are
localized (instead of being set by $g_5 \sqrt{k}$). As the size of
the brane kinetic terms increases, it turns out that the gauge KK
coupling (measured in units of the SM gauge coupling as usual)
becomes weaker \cite{Davoudiasl:2002ua}. At the same time, the mass of
the lightest KK mode becomes smaller in such a way that ratio
\begin{eqnarray}
\frac{\text{KK coupling constant}}{\text{Lightest KK mass} \;
\hbox{in units of} \; k e^{ - k \pi R } }
\end{eqnarray}
stays roughly the same (for moderately large brane terms), up to
$O(1)$ factors.
The flavor-violating amplitude (in units of $k e^{ - k \pi R }$)
depends on precisely the above ratio.
So it is clear that large TeV brane terms can allow lighter KK
states to satisfy the flavor constraints, but it will {\em not} allow a
reduction in the scale $k e^{- k\pi R}$ which might be the one more relevant
(than the lightest KK mass scale) for the fine tuning in EWSB.
Although a detailed analysis of TeV brane kinetic terms is beyond
the scope of this paper, it is important to keep in mind that such
terms can affect the bounds on the scale $k e^{ -k \pi R }$ by
$O(1)$ factors.
Finally, for smaller than Planck-weak hierarchy,
for example as in the ``Little Randall-Sundrum model''
\cite{Davoudiasl:2008hx},
it is clear that $g_5\sqrt{k}$ can be smaller as seen from Eq.
(\ref{match}).

\subsubsection{Perturbativity bound on size of $5D$ gauge coupling}
On the other hand, an {\em upper} bound on $g_5^{ QCD }$ coupling can also be obtained from
the condition of perturbativity of the $5D$ QCD theory in the
following way. We can estimate the loop expansion parameter for this
theory by comparing the one-loop correction to the tree-level value
of a coupling (or comparing a two-loop correction to a one-loop
effect). This loop expansion parameter grows with energy (or number
of active KK modes) due to the non-renormalizability of $5D$
couplings. So, the number of KK modes below the $5D$ cut-off,
denoted by $N_{ KK }$, can then be estimated by setting this loop
expansion parameter to be $\sim 1$ (see, for example
\cite{Appelquist:2000nn}). As an example, we can {\em estimate} the
one-loop correction to the tree-level value of the three KK gluon
coupling arising from this interaction itself. Including color and
helicity factors of $\sim 3$ each for this loop diagram (see, for
example, reference \cite{Choi:2002ps}), we find:
\begin{eqnarray}\label{gqcd}
\label{gluon pert} \frac{ \left( g^{ QCD }_5 \sqrt{k} \right)^2 3
\times 3 }{ 16 \pi^2 } N_{ KK } & \sim & 1 \label{NKK}
\end{eqnarray}
where $\sim \left( g_5^{QCD} \sqrt{k} \right)$ is coupling of $3$ KK
gluons and the single power of $N_{ KK }$ (i.e., single KK sum)
follows from KK number conservation at the purely KK gluon vertices.
Equivalently, the dimension of $g_5^{QCD}$ being $-1/2$ implies that
the $5D$ loop expansion parameter is $\sim (g_5^{\text QCD})^2 E /
\left( 16 \pi^2 \right)$ with $E / k \sim$ being the number of
active KK modes.

We can also instead consider the one-loop self-corrections to the
coupling of KK gluon to two KK fermions, where the helicity factor
of $3$ is absent (in this sense, the estimate in Eq.(\ref{gqcd}) is
conservative). The estimate in Eq. (\ref{NKK}) leads to the
following values of the number of KK modes below cutoff:
\begin{itemize}

\item
$N_{ KK } \sim 2$ for $g_5^{QCD} \sqrt{k} \sim 3$ which is again the smallest
$g_5^{ QCD } \sqrt{k}$ allowed for Planck-weak hierarchy (i.e., with
loop-level matching of the $5D$ coupling to the $4D$ coupling and
with no bare/tree-level brane kinetic terms).

\item
Whereas for $g_5^{QCD} \sqrt{k} \sim 6$ (i.e., with tree-level matching of
the $5D$ coupling to the $4D$ coupling with no brane kinetic terms),
there seems to be hardly any energy regime where the $5D$
theory is weakly coupled, i.e., $N_{ KK } < 2$.

\end{itemize}
This conclusion about perturbativity for the $g_5^{QCD} \sqrt{k}
\sim 6$ case is valid even if we do {\em not} include the helicity factor
of $\sim 3$ as would be the case for the estimate of loop expansion
parameter using the KK gluon coupling to two KK fermions (instead of
coupling of three KK gluon coupling). So with this perturbativity
motivation (and using the correspondence in Eq.
\ref{relationgauge}), we have
%
%
focused on
using $g_{ s \ast } \sim 3$ in
our analysis of the two-site model, but of course, one should
understand that these conclusions are just estimates.

\subsection{Formulae for zero-mode and KK Yukawa and KK gauge couplings}\label{KKcoupling}

We give some useful formulae for profiles of zero-mode fermion, KK
fermion, gauge KK mode and bulk Higgs in $5D$ AdS model, neglecting
(for simplicity) brane kinetic terms\footnote{Since the
KK modes are localized near the TeV brane, localized
kinetic terms there
affect the KK decomposition and generate additional 
flavor-violating
couplings of gauge KK modes to fermion zero-modes (see,
for example, reference \cite{Csaki:2008eh}).
However, we assume
that these brane terms
are of size generated by loop-level bulk effects (which is a
technically natural choice) so that 
these effects can be neglected for our 
purposes,
as long as the bulk loops are perturbative.}
(see, for example, reference
\cite{Agashe:2004cp} for fermion and gauge profiles and
\cite{Cacciapaglia:2006mz} for Higgs profile).

First, we decompose the 5D fermion field as
\begin{eqnarray}
\Psi(x, z) = \sum_n \psi^{(n)}(x)\chi_n(c,z)
\end{eqnarray}
where
$c$ is the ratio between $5D$ fermion mass term and AdS curvature scale
$k$. The normalization condition for profiles is
\begin{eqnarray}
\int_{z_h}^{z_v} dz \left(\frac{z_h}{z} \right)^4 \chi_n^2(c,z) =
1
\end{eqnarray}
The fermion zero mode profile is
\begin{equation}
\chi_0 (c, z) = f(c) \left(\frac{z}{z_h}\right)^{2-c}
\frac{1}{\sqrt{z_h}}\left(\frac{z_h}{z_v}\right)^{1/2-c}
\end{equation}
with
\begin{equation}
f(c) = \sqrt{\frac{1-2c}{1-(z_v/z_h)^{2c-1}}}
\end{equation}
The KK fermion profile {\em for the same chirality
as the zero-mode}
%
%
is
\begin{equation}
\chi_n(c,z) = \left(\frac{z}{z_h}\right)^{5/2} \frac{1}{N^\chi_n
\sqrt{\pi r_c}} [J_\alpha(m_n z) + b_\alpha(m_n) Y_\alpha(m_n z)]
\label{KKfermion}
\end{equation}
with $\alpha = |c+1/2|$, $m_n$ and $b_\alpha$ are given by
\begin{equation}
\frac{J_{\alpha\mp 1} (m_n z_h)}{Y_{\alpha\mp 1}(m_n z_h)} =
\frac{J_{\alpha\mp 1} (m_n z_v)}{Y_{\alpha\mp 1}(m_n z_v)} =
-b_\alpha(m_n)
\end{equation}
with upper (lower) signs for $c > -1/2 \;(c < -1/2)$ and
normalization condition gives
\begin{equation}
|N^\chi_n|^2 = \frac{1}{2\pi r_c z_h} \left[ z_v^2[J_\alpha(m_n z_v)
+ b_\alpha(m_n) Y_\alpha(m_n z_v)]^2 - z_h^2[J_\alpha(m_n z_h) +
b_\alpha(m_n) Y_\alpha(m_n z_h)]^2 \right]
\end{equation}
It is useful to note that the ratio of zero and KK fermion profile
at TeV brane is
\begin{eqnarray}\label{zerokkprofile}
\frac{\chi_0(c, z_v)}{\chi_n(c, z_v)} \approx \frac{f(c)}{\sqrt{2}}
\end{eqnarray}
Similarly, we perform KK decomposition for gauge bosons:
\begin{eqnarray}
{\cal A}_\mu(x,z) = \sum_n A^{(n)}(x) f_n(z)
\end{eqnarray}
The gauge KK wavefunction is:
\begin{equation}
f_n(z) = \sqrt{\frac{1}{z_h}} \frac{z}{N^f_n} [J_1 (m_n z) + b_n
Y_1(m_n z)]
\end{equation}
where $b_n$ and gauge KK masses are fixed by:
\begin{eqnarray}
\frac{J_0(m_n z_h)}{Y_0(m_n z_h)} = \frac{J_0(m_n z_v)}{Y_0(m_n
z_v)} = - b_n
\end{eqnarray}
and the normalization condition
\begin{equation}
\int_{z_h}^{z_v} dz \left(\frac{z_h}{z} \right) f_n^2(z) = 1
\end{equation}
gives us
\begin{equation}
|N_n^f|^2 = \frac{1}{2}\left[ z_v^2[J_1(m_n z_v) + b_n Y_1(m_n
z_v)]^2 - z_h^2[J_1(m_n z_h) + b_n Y_1(m_n z_h)]^2 \right]
\end{equation}

The KK decomposition for a $5D$ scalar (bulk Higgs)
is
(here $\beta = \sqrt{4+\mu^2}$, with $\mu$ being the bulk Higgs mass in
units of $k$):
\begin{equation} {\cal H} (x, z) = v( \beta, z) + \sum_n H^{(n)}(x) \phi_n(z)
\end{equation}
where $v( \beta, z)$ is the Higgs {\em vev} profile, which is very close to
the
(lightest) {\em physical} Higgs profile when $m_h \ll M_{KK}$. This profile
can be chosen to be peaked near the TeV
brane:
\begin{equation}\label{bulkHiggs}
v( \beta, z) = v_4 z_v
\sqrt{\frac{2(1+\beta)}{z_h^3(1-(z_h/z_v)^{2+2\beta})}}
\left(\frac{z}{z_v}\right)^{2+\beta}.
\end{equation}

The couplings between fermion zero modes and Higgs ($Y_0$),
fermion KK modes and Higgs ($Y_{KK}$), fermion zero modes and
gauge KK modes ($g_{KK}$) are given by overlap integrals of the
their profiles multiplied by the $5D$ couplings:
\begin{eqnarray}
Y_0 \left( c_L, c_R, \beta \right) & = & Y^{ bulk }_5 \int dz
%
%
\left( \frac{z_h}{z}\right)^5 \; v(\beta, z) \chi_{0L}(c_L, z)
\chi_{0R}(c_R, z)/v_4
\nonumber \\
%
%
Y_{ KK } \left( c_L, c_R, \beta
\right) & = &Y^{ bulk }_5 \int dz
%
%
\left( \frac{z_h}{z}\right)^5 \;v(\beta, z) \chi_{nL}(c_L, z)
\chi_{mR}(c_R, z)/v_4\nonumber \\
g_{ KK }
\left( c_L
\right) & = & g_5\int dz
%
%
\left( \frac{z_h}{z}\right)^4 \;f_n(z) \chi_{0L}(c_L, z)
\chi_{0L}(c_L, z)
%
%
\end{eqnarray}
where $Y_5^{ bulk }$ is defined by ${\cal S } \ni \int d^4 x dz
\sqrt{G}\;Y_5^{ bulk } H( x, z ) \overline{ \Psi ( x, z ) } \Psi^{
\prime } ( x, z )$ (with $\Psi$ and $\Psi^{ \prime }$ being
$SU(2)_L$ doublet/singlet
and $G$ is the determinant of the metric) and has mass dimension $-1/2$ just like
$g_5$. Again, $Y_{ KK }$ defined above is for KK modes with
same chirality
as the zero-mode.
%
%
A similar expression can be obtained for the overlap
integrals giving the
coupling between KK gluon and two KK fermions which was used
to obtain Eq. (\ref{relationgauge}).

It is useful to know approximate formulae for these overlap
integrals \cite{Agashe:2004cp}\cite{Csaki:2008zd}. For example
\begin{eqnarray}\label{bulkgaugecoupling}
g_{ KK } & \approx &  \left( g_5 \sqrt{k} \right)\left(-\frac{1}{k
\pi r_c} + f(c_L) f(c_R)\right)
\end{eqnarray}
where pre-factor of ``$1$'' that multiplies $f(c_L) f(c_R)$ is
almost $c$-independent for $0.4 \stackrel{<}{\sim} c
\stackrel{<}{\sim} 0.7$ that is of interest for down-type quarks.

Similarly, we define the parameter $a ( \beta, c_L, c_R )$
%
%
by
\begin{eqnarray}\label{a}
Y_0 \left( c_L, c_R, \beta \right) & = & a
( \beta, c_L, c_R ) Y_{ KK } \left( c_L, c_R, \beta \right)
f(c_L) f(c_R)
%
%
\end{eqnarray}
%
%
%
We find (numerically) that, for fixed Higgs vev profile, the
$c_{ L, R }$ dependence of $a$ is {\em very} mild for the range
$0.4 \stackrel{<}{\sim} c
\stackrel{<}{\sim} 0.7$ that is of interest for the down-type quarks
and hence we set $c_L = c_R = 0.55$ henceforth when we quote values
of $a$.
We give a table for $a$ vs. the parameter $\beta$ of bulk Higgs (see
Table \ref{table}).
\begin{table}[hbt]
\begin{center}
\begin{tabular}{|c|c|c|c|}
\hline $\beta$ & $a$  & $M_{KK}$
%
%
\small{($g_{5}^{QCD}\sqrt{k} = 3$, $Y_{KK} = 6$)}  & $M_{KK}$
%
%
\small{($g_{5}^{QCD}\sqrt{k} = 6$, $Y_{KK} = 6$)}  \\
\hline
0 & 1.5 & 3.7 TeV & 7.4 TeV\\
\hline
1(two-site) & 1 & 5.5 TeV & 11 TeV\\
\hline
2 & 0.75
& 7.3 TeV & 14.6 TeV\\
\hline
$\infty$ (brane) & 0.5 & 11 TeV & 22 TeV\\
\hline

%
%
%
\end{tabular}
\end{center}

\caption[]{\label{table} The values of the parameter $a$
(relating zero
to KK mode Yukawa couplings: see Eq. (\ref{a})) in 1st column
for different values
of the parameter $\beta$ (2nd column) which determines the
profile of the bulk Higgs (Eq. (\ref{bulkHiggs})).
The two-site model and brane Higgs case are also shown as
corresponding to
specific values of $\beta$ (see discussion in text). The
bound on $M_{KK}$ (from $\epsilon_K$ {\em only},
based on the estimate in Eq. (\ref{KKbound5D})) for the purely composite
sector (or KK) gauge coupling $g_{5}^{QCD}\sqrt{k} = 3$
(3rd column) and
$g_{5}^{QCD}\sqrt{k} =6$ (last column)
are also shown. We fix the composite/KK Yukawa coupling
$Y_{KK} = 6$ for all entries in the table and $c_L = c_R = 0.55$ in order to obtain the
value of $a$.}

\end{table}
We see that $a \sim O(1)$ as expected.
In detail,
the Higgs and KK fermion profiles are localized near the TeV brane
so that
$Y_{ KK }$ is dominated by overlap of profiles in this region.
So, we get $Y_{ KK } \sim Y_5 \sqrt{k}$ (with a mild dependence
on $c$ and $\beta$),
where the $5D$ Yukawa is made dimensionless simply by
a factor of $\sim \sqrt{k}$ coming from the normalized
profiles
at the TeV brane: see Eqs. (\ref{KKfermion}) and (\ref{bulkHiggs}).
Even though the fermion zero-modes (except for top quark) are
localized near the Planck brane, their overlap with the Higgs is
still dominated by the region near the TeV brane for the choices of
$c$'s relevant for quark masses\footnote{For larger values of $c$'s
(i.e., fermion zero-modes localized closer to the Planck
brane)
as relevant for Dirac neutrino masses, the overlaps with Higgs can be dominated
by the region near the Planck brane instead \cite{Agashe:2008fe}.}.
Therefore, using the ratio of fermion zero and KK mode profiles
($f$'s) given in Eq. (\ref{zerokkprofile}), we expect $Y_0 \sim Y_{
KK } f \left( c_L \right) f \left( c_R \right) \sim \left( Y_5
\sqrt{k} \right) f \left( c_L \right) f \left( c_R \right)$, i.e.,
$a \sim O(1)$.
Note that $f \left( c \right)$'s can be hierarchical even with small
variations in $c$'s,
resulting in a
solution to the flavor hierarchy problem in the sense that
$4D$ Yukawa matrix ($Y_0$) can be hierarchical without any (large) hierarchies
in the $5D$ theory, i.e., with anarchic $5D$ Yukawa
matrix (or $Y_{ KK }$) and $O(1)$ $c$'s.

The following
observation about the parameter $a$ is crucial
for the analysis of $\epsilon_K$ in next section.
Since the fermion
zero modes profiles peak near the Planck brane while the fermion KK mode
profiles
peak near the TeV brane, it is clear that the overlaps of profiles
of fermion zero modes with Higgs
increase while those of fermion KK modes with Higgs
decrease as the Higgs wavefunction moves farther {\em away} from the TeV
brane. Therefore,
as seen from this table,
\begin{itemize}

\item
as we {\em decrease} the parameter $\beta$ determining the
Higgs profile in Eq. (\ref{bulkHiggs}) -- thereby
localizing the Higgs away from the TeV brane, the
parameter $a$ in Eq. (\ref{a}) {\em increases}.

\end{itemize}

We thus expect the opposite limit, $\beta \rightarrow \infty$, to reproduce brane
Higgs scenario. In fact,
for brane-localized Higgs, couplings of fermions to Higgs are simply given
by wavefunctions of fermions at TeV brane, i.e.,
there is {\em no} overlap integral to be performed:
\begin{eqnarray}\label{brane}
Y^{ brane }_0 & = & \left( Y^{ brane }_5 k \right) f_L f_R
\nonumber \\
%
%
Y^{ brane }_{ KK } & = & \left( 2 Y^{ brane } _5 k \right)
\end{eqnarray}
with ${\cal S} \ni \int d^4 x \sqrt{G} Y_5^{ brane } H (x)  \overline{ \Psi_L
( x, z_{ v } ) } \Psi^{ \prime }_R ( x, z_{ v } )$. Note that
dimension of $Y_5$ changes from $-1/2$ to $-1$ as we switch from
bulk Higgs to brane-localized Higgs. The factor of two in $Y^{ brane
}_{ KK } $ in second line of Eq. (\ref{brane}) comes from the fact
that the normalized KK wavefunction at TeV brane is $\approx \sqrt{
2 k }$ (see Eq. (\ref{KKfermion})). From Eqs. (\ref{a}) and
(\ref{brane}), the model with brane-localized Higgs (effectively)
has $a = 1/2$. And, the numerical calculation of the overlap
integrals for bulk Higgs shows that indeed 
$a \rightarrow 1/2$ for $\beta
\rightarrow \infty$ (see Table \ref{table}), in agreement with the
above expectation.

Now we can see the similarity between the two-site model and the bulk Higgs
scenario. First, we compare the gauge couplings between the two
cases:  Eq. (\ref{bulkgaugecoupling}) and ${\cal{L}}^{SM-SM}$ term
of Eq. (\ref{gauge}), using Eq. (\ref{relationgauge}).
From these equations, we can make the
following identifications:
\begin{eqnarray}
 s_{L,R} &\leftrightarrow  & f_{L,R} \\ \nonumber
\frac{1}{k\pi
r_c}& \leftrightarrow & \tan^2\theta
\end{eqnarray}
As mentioned above, $f_{L_i,R_i}$ can be hierarchical with small variations in
$5D$ fermion mass parameters (c). Therefore, our choice of
hierarchical elementary/composite mixing angles ($s_{q,u,d}$)
in the two-site model
%
%
is justified.

We turn to Yukawa couplings and compare Eq. (\ref{a}) with ${\cal{L}}
_{Y}^{\text{SM-SM}}$ term of Eq. (\ref{yukawa}). First, just like
for the gauge couplings, we should identify the
Higgs coupling to heavy fermions in the
two-site model with the Higgs coupling to KK fermions in the
$5D$ model\footnote{Note that, for a fixed $\beta$, $Y_{ KK }$
is only mildly sensitive to $c_{ L, \; R }$'s.}, i.e.,
\begin{equation}
Y_\ast \leftrightarrow Y_{KK}
\end{equation}
(In particular, both are assumed to be anarchic.)
Then we can see that the two-site and
$5D$ Yukawa coupling equations match if $a
= 1$. Therefore, we conclude that
\begin{itemize}
\item the two-site model ``mimics'' the bulk Higgs scenario with $\beta \approx
1$ (which has $a \approx 1$).
This result is also shown in Table \ref{table}.

\end{itemize}

\subsection{Bound from $\epsilon_K$}

Following the arguments of the analysis of $\epsilon_K$ for the
two-site model, it is clear that, in the bulk Higgs scenario, we get
from KK gluon exchange
\begin{eqnarray}
C^{ 5D }_{ 4 \; \hbox{estimate} } \left( M_{ KK } \right) & = &  \frac{
\left( g_5 \sqrt{k} \right)^2 }{ Y_{ KK }^2 a \left)( \beta \right)^2 }
\frac{ 2 m_s m_d }{
v^2 } \frac{1}{ M_{ KK }^2 },
\end{eqnarray}
where ``estimate'' has the same meaning as in our
analysis of the two-site model.
Thus the constraint from $\epsilon_K$ is
\begin{eqnarray}
M_{ KK } \stackrel{>}{\sim}11 \frac{ g_5 \sqrt{k} }{ Y_{ KK } a ( \beta )
}\hbox{TeV} \label{KKbound5D}
\end{eqnarray}
The bounds on $M_{KK}$ for different values of $\beta$ (i.e.,
choices of Higgs profiles), including the brane Higgs case and the
two-site model is shown in Table \ref{table} for $g_{5}^{QCD} =
3, 6$ and $Y_{ KK } = 6$.

Now we can compare our results to previous analysis: references
\cite{Csaki:2008zd, Blanke:2008zb} used a brane-localized Higgs,
i.e., $a \sim 1/2$, with $Y^{ brane }_5 k \sim 3$, i.e., $Y_{ KK }
\sim 6$ (from Eq. \ref{brane}).
%
%
They obtained the bound on KK scale of
$\sim 20 (10)$ TeV for the case of $g_5^{QCD} \sqrt{k} \sim 6
(3)$
which agrees with
our results in Table \ref{table}.
However, from Table
\ref{table}, we see that
\begin{itemize}

\item
for {\em same} $g_5 \sqrt{k}$ and KK Yukawa ($Y_{ KK }$), the bound
on $M_{ KK }$ from $\epsilon_K$ is lowered for a {\em bulk} Higgs (instead
of brane-localized Higgs).

\end{itemize}
Of course, this reduction in the KK scale for a bulk Higgs relative
to the case of brane localized Higgs is due to a smaller coupling of
SM fermions to the KK gluon for the bulk Higgs case, i.e.,
the zero-mode fermions being localized
a bit farther from the TeV brane (where gauge
KK modes are localized), than for the brane-localized Higgs case.
The crucial point is that, even with this shift of zero-mode fermion
profiles relative to the brane-localized Higgs case, the bulk Higgs
set-up can maintain the {\em same} (i.e., SM) value of the zero-mode
Yukawa (for the {\em same} KK Yukawa) as in brane-localized Higgs case.
Here, we use the result (explained above) that the ratio of zero-mode to
KK Yukawa couplings (denoted by $a$ above) is larger for the bulk Higgs case
than for brane-localized Higgs (for fixed fermion profiles).

We remind the reader that we are {\em not} considering models where
Higgs is the $5^{ \hbox{th} }$ component of $5D$ gauge field here.
In the Higgs-as-$A_5$ model, the SM Higgs also has a profile which
is peaked near the TeV brane {\em in a specific gauge}
\cite{Contino:2003ve}. However, for this model, it was shown in
reference \cite{Csaki:2008zd} that the lower limit on the KK mass
scale is $\sim 10$ TeV for the choices $g^{ QCD }_5 \sqrt{k} \sim 3$
and $g^{ EW }_5 \sqrt{k} \; (\hbox{which is the ``effective'' $5D$
Yukawa}) \sim 6$. For larger $g^{ QCD }_5 \sqrt{k}$ and/or smaller
$g^{ EW }_5 \sqrt{k}$, the bound on KK scale is higher.

\subsection{Perturbativity limit on size of KK
Yukawa}\label{perturbativity}
%
%
Finally, we wish to illustrate why $\epsilon_K$ {\em by itself} might allow
{\em a few}, say, $\sim 3$ TeV KK scale, even with {\em anarchy}
in $5D$ flavor parameters, i.e.,
mixing angles of size as in Eq. (\ref{mixing angles}).
The point is that the bound on KK scale from $\epsilon_K$ depends on
size of KK Yukawa as seen in Eq. (\ref{KKbound5D}).
Instead of using $b \rightarrow s \gamma$ in order to constrain $Y_{
KK }$ (as we did for the two-site model), we can use perturbativity of the $5D$
theory.

Proceeding in the same way as for the
%
%
gluon coupling, we can estimate $N_{ KK }$ from the loop expansion
parameter associated with the Yukawa coupling being $\sim 1$. For
example, we can compare the one-loop correction to the tree-level
value of the coupling of Higgs to two
%
%
KK fermions from this coupling itself (there are no color or
helicity factors here). For {\em brane}-localized Higgs, we get
\begin{eqnarray}
\frac{ Y^{ brane \; 2 }_{ KK } }{ 16 \pi^2 } N_{ KK }^2 & \sim & 1
\end{eqnarray}
where $N_{ KK }^2$ (i.e., {\em double} KK sum) in this loop diagram
follows from
absence of KK number conservation at the Higgs vertices in the
brane-localized Higgs case. One can also derive such growth of the loop
expansion parameter with $N_{ KK }$ from dimensional analysis,
namely,
%
%
$[Y^{ brane }_5]=-1$ such that the $5D$ loop expansion parameter is $\sim
Y^{ brane \; 2 }_5 E^2 / \left( 16 \pi^2 \right)$. So, for the brane-localized
Higgs case, we get $Y^{ brane }_{ KK } \sim 4 \pi / N_{ KK }$ and
the choice of $Y_{ KK } \sim 6$ (i.e., $Y^{ brane }_5 k \sim 3$) in
references \cite{Csaki:2008zd, Blanke:2008zb} for brane Higgs
corresponds to $N_{ KK } \sim 2$.

On the other hand, the loop expansion parameter for the {\em bulk}
Higgs case is
\begin{eqnarray}
\frac{ Y^{ bulk \; 2 }_{ KK } }{ 16 \pi^2 } N_{ KK } & \sim & 1
\end{eqnarray}
where the {\em single} power of $N_{ KK }$ follows from the single KK sum
due to KK number conservation at Higgs vertices for the bulk Higgs
case. Equivalently, we can use dimensional analysis, i.e.,
$[Y^{ bulk }_5]=-1/2$ so that the $5D$ loop expansion parameter $\sim
Y^{ bulk \; 2 }_5 E
/ \left( 16 \pi^2 \right)$ just like for $5D$ gauge theory. Hence,
we have for bulk Higgs case, $Y^{ bulk }_{ KK } \sim 4 \pi / \sqrt{
N_{ KK } }$, i.e.,

\begin{itemize}

\item
for same $N_{ KK }$, we find that $Y_{ KK }$ can be larger for bulk
Higgs by $\sim \sqrt{ N_{ KK } }$ than for the brane-localized Higgs
case. Thus the KK mass bound can be lowered even further ({\em beyond} the
point related to the factor $a$ discussed above) as seen from Eq
(\ref{KKbound5D}): see also discussion in \cite{Fitzpatrick:2007sa}.

\end{itemize}
And, in particular,
\begin{itemize}

\item
we get $Y^{ bulk }_{ KK } \sim 6 \sqrt{2}$ for $N_{ KK } \sim 2$
(same as the choice made in references \cite{Csaki:2008zd, Blanke:2008zb}) so
that choosing in addition
%
%
the Higgs profile with $\beta \sim 0$ (so that $a \sim 1.5$) and
$g_{ \ast } \sim 3$, we see from Eq. ({\ref{KKbound5D})} that $M_{
KK } \sim 2.6$ TeV might be allowed by $\epsilon_K$
constraint.

\end{itemize}

However, such a low KK scale and large $Y_{ KK }$
in the $5D$ model will most likely be very
strongly
constrained
%
%
by $\hbox{BR}\left( b \rightarrow s \gamma
\right)$ just as in the case of the two-site model. Note that the bulk
Higgs couplings other than $Y_{ 0, \; KK }$ --
for example the mixed (i.e., zero-KK fermion)
ones -- might not {\em exactly} mimic the corresponding ones in the two-site
model so that our results for $b \rightarrow s \gamma$ in the
two-site model cannot be directly used for the $5D$ model\footnote{Of
course, the amplitude for $b \rightarrow \gamma$
in the $5D$ model is expected to be of similar size to (i.e., differing only
by $\sim O(1)$ factors from) 
that in the two-site model.}.
A
detailed calculation of $b \rightarrow s \gamma$ for the $5D$ model is
beyond the scope of this work.


\newpage


\end{document}